\documentclass[lineno]{jfm}

\usepackage{graphicx}
\usepackage{newtxtext}
\usepackage{newtxmath}
\usepackage{natbib}
\usepackage{tikz}
\usetikzlibrary{quotes,arrows.meta,matrix}
\usetikzlibrary{math}
\usetikzlibrary{positioning}
\usepackage{soul}
\usepackage{pgfplots}
\definecolor{kit-red100}{rgb}{.635,.133,.137}
\definecolor{kit-green100}{rgb}{0,.59,.51}
\definecolor{kit-blue100}{rgb}{0.275,0.392,0.667}
\pgfplotsset{compat = 1.3}
\usepackage[outline]{contour}
\usepgfplotslibrary{fillbetween,groupplots}
\usepackage{bm}  
\usepackage{subcaption}
\usepackage{hyperref}
\hypersetup{
    colorlinks = true,
    urlcolor   = blue,
    citecolor  = black,
}

\newcommand{\RomanNumeralCaps}[1]
\linenumbers
\pgfplotsset{compat=1.15
 ,colormap={parula}{
rgb255=(53,42,135)
rgb255=(15,92,221)
rgb255=(18,125,216)
rgb255=(7,156,207)
rgb255=(21,177,180)
rgb255=(89,189,140)
rgb255=(165,190,107)
rgb255=(225,185,82)
rgb255=(252,206,46)
rgb255=(249,251,14)
        }}

\title{Prediction of equivalent sand-grain size and identification of drag-relevant scales of roughness -- a data-driven approach}

\author{Jiasheng Yang\aff{1},
  Alexander Stroh\aff{1},
  Sangseung Lee\aff{2,3},
  Shervin Bagheri\aff{4}, Bettina Frohnapfel\aff{1}
 \and Pourya Forooghi\aff{5}
 \corresp{\email{\href{mailto:forooghi@mpe.au.dk}{forooghi@mpe.au.dk}}}}

\affiliation{\aff{1}Institute of Fluid Mechanics, Karlsruhe Institute of Technology, 76131 Karlsruhe, Germany
\aff{2} Department of Mechanical Engineering, Inha University, Incheon 22212, Republic of Korea
\aff{3} Applied AI Center for Thermal and Fluid Research, Inha University, Incheon 22212, Republic of Korea
\aff{4} Department of Engineering Mechanics, FLOW centre, KTH, Stockholm SE-100 44, Sweden
\aff{5} Department of Mechanical \& Production Engineering, Aarhus University, 8200 Aarhus C, Denmark}

\begin{document}
\maketitle

\begin{abstract}
Despite decades of research, a universal method for prediction of roughness-induced skin-friction in a turbulent flow over an arbitrary rough surface is still elusive.
The purpose of the present work is to examine two possibilities; firstly, predicting equivalent sand-grain roughness size $k_s$ based on the roughness height probability density function (p.d.f.) and power spectrum (PS) leveraging machine learning as a regression tool, and secondly, extracting information about relevance of different roughness scales to skin-friction drag by interpreting the output of the trained data-driven model. The model is an ensemble neural network (ENN) consisting of 50 deep neural networks. The data for the training of the model is obtained from direct numerical simulations (DNSs) of turbulent flow in plane channels over 85 irregular multi-scale roughness samples at friction Reynolds number Re$_\tau=800$.
The 85 roughness samples are selected from a repository of 4200 samples, covering a wide parameter space, through an active learning (AL) framework. 
The selection is made in several iterations, based on the informativeness of samples in the repository, quantified by the variance of ENN predictions. This AL framework aims to maximize the generalizability of the predictions with a certain amount of data. This is examined using three different testing data sets with different types of roughness, including 21 surfaces from the literature. The model yields an overall mean error of 5\% to 10\% on different testing data sets. 
Subsequently, a data interpretation technique, known as layer-wise relevance propagation, is applied to measure the contributions of different roughness wave-lengths to the predicted $k_s$.
High-pass filtering is then applied to the roughness PS to exclude the wave-numbers identified as drag-irrelevant.
The filtered rough surfaces are investigated using DNS, and it is demonstrated that, despite significant impact of filtering on the roughness topographical appearance and statistics, the skin-friction coefficient of the original roughness is successfully preserved.
\end{abstract}



\section{Introduction}
\label{sec:intro}
Surface degradation in flow-related engineering applications can take various forms, such as wearing or fouling, resulting in roughness on the solid surfaces. The most significant effect of surface roughness, in practical sense, is an increase in the skin-friction drag under turbulent flow conditions. As an example, the uncertainties in the prediction of roughness-induced skin-friction on ship hulls subjected to bio-fouling can cause multiple billion dollars of energy waste every year~\citep{Schultz2011,Chung2021annrev}.
Understandably, study of turbulent flow over rough surfaces has been an active area of research for nearly a century \citep{Nikuradse1933,schlichting1936experimentelle,perry_schofield_joubert_1969,Raupach1992,krogstad1992,bhaganagar2004,busse_thakkar_sandham_2017,aghaei2022}.

The seminal work by \citet{Nikuradse1933} has provided the following researchers with a common `currency' to measure roughness-induced drag; that is equivalent sand-grain roughness size, $k_s$, defined as the sand-grain size in Nikuradse's experiments producing the same skin-friction coefficient as a rough surface of interest in the fully-rough regime.
Equivalent sand-grain size is related to the downward shift in the logarithmic region of the inner-scaled mean velocity profile widely observed on rough walls, which is referred to as roughness function $\Delta U^+$~\citep{hama1954}. In the fully-rough regime, where the skin-friction coefficient is independent of Reynolds number,
\begin{equation}
\label{Eqn_EquivalentSand}
    \Delta U^+=\frac{1}{\kappa}\text{ln}(k_s^+)+B-8.5~,
\end{equation}
where $\kappa$ is the von K\'arm\'an constant and $B$ is the smooth-wall log-law intercept~\citep{jimenez04}. 

One must note that $k_s$ is, by definition, a flow variable and not a geometric one. As a result, for any `new' rough surface, it needs to be determined through a (physical or high-fidelity numerical) experiment in which the skin-friction drag is measured. Obviously, such an exercise is not practical in many applications; therefore, a great amount of effort in the past few decades has been devoted to determining $k_s$ of an arbitrary roughness \textit{a priori}, i.e. merely based on its geometry (see e.g. \citet{vanrij02,flack2010,chan2015,10.1115/1.4037280,Thakkar2017,Flack2020}). A comprehensive description of these efforts can be found in the reviews by \citet{Chung2021annrev} and \citet{flack2022}.
Essentially, they can be summarized as attempts to regress correlations between $k_s$ (or $\Delta U^+$) and a few statistical parameters of roughness geometry based on available data. Some widely used parameters in this context are skewness of roughness height probability density function (p.d.f.) \citep{flack2010}, effective (or mean absolute) slope \citep{napoli_armenio_demarchis_2008}, and correlation length of rough surface geometry \citep{Thakkar2017}.

As a result of increased computational capacities in recent years, DNS has become a source of data for development of accurate roughness correlations as pointed out by \citet{flack18}. In this regard, the idea of DNS in minimal channels, proposed by \citet{chung_chan_macdonald_hutchins_ooi_2015}, has enabled characterizing larger numbers of roughness samples with a certain computational resource. Availability of more data, on one hand, has opened the door to utilization of machine learning (ML)-based regression tools, and on the other hand, enables inclusion of more roughness information (beyond only a few parameters) as the input to such tools. The latter point is particularly important since there is increasing evidence that both statistical and spectral information on roughness geometry are required for prediction of flow response to a multi-scale roughness \citep{Alvesportela21}. In this regard, it has been shown that $k_s$ for multi-scale random roughness can be determined nearly uniquely with a combined knowledge of roughness height p.d.f. and its power spectrum (PS)  \citep{yang_stroh_chung_forooghi_2022,IHFF_Yang_2022}.

The first ML-based `data-driven' tool for prediction of $k_s$ has been recently reported by \citet{jouybari_2021}. These authors used deep neural network and Gaussian process regression to train models with 17 inputs including widely used roughness parameters and their products. The training data for their model is obtained from DNS of flow over certain types of artificially generated roughness, which were also used to evaluate the model. \citet{lee_yang_forooghi_stroh_bagheri_2022} used a similar {\color{black}neural network} to that of \citet{jouybari_2021} and showed that improvements in predictive performance can be achieved if the network is `pre-trained' on existing empirical correlations. While these pioneering works deliver promising results, the data-driven approach arguably has the potential to realize truly universal models, which can generalize beyond a certain class of roughness. The present work is an attempt to explore this potential. To this end a model is trained on a wide variety of multi-scale irregular roughness samples, selected based on an adaptive approach (explained shortly), which is aimed to enhance the universality of the predictions. This is evaluated using `unseen' roughness from  different testing data sets with different natures. Moreover, unlike the previous efforts, the present model incorporates the complete p.d.f. and PS of roughness as inputs rather than a finite set of pre-determined parameters.

Considerable attention has been paid in recent literature to the multi-scale nature of realistic roughness and the significance of its `spectral content'. 
It has been suggested that, beyond a certain threshold, large roughness wave-lengths may impact the roughness-induced drag less significantly \citep{BARROS20181,yang_stroh_chung_forooghi_2022}. While parametric studies of roughness PS \citep{anderson_meneveau_2011,BARROS20181} or Fourier filtering \citep{BUSSE2015129,Alvesportela21} can shed light on this matter, in the present work we explore the possibility to directly evaluate contributions of different roughness scales utilizing the information embeded in the data-driven model developed in this study. This is motivated by the fact that the (discretized) PS is a direct input to the model, which hints at the potential to extract information about the role of different wave-lengths through interpretation of the model.

In order to train the data-driven roughness model, $k_s$ for several roughness samples should be determined. This is referred to as `labeling' of those samples borrowing the term from the ML terminology. Moreover, each roughness sample along with its $k_s$ value is called a training `data point'. One should note that labeling is a computationally expensive process due to the need to perform DNS.
In dealing with such scenarios, ML methods classified under active learning (AL) -- also known as query-based learning~\citep{Abe1998QueryLS} or optimal experimental design~\citep{fedorovtheory} in different contexts -- have been proven particularly advantageous  ~\citep{zhu05c_interspeech,Settlesetal2008,Bangert2021}. 
In AL, selection of the training data is navigated in a way that the information gain from a certain amount of available data is maximized~\citep{Settles09}.
The `informativeness' of a potential data point is commonly measured by the uncertainty in its prediction, which needs to be determined without labeling, e.g. through the standard deviation of the predictive distribution of a Bayesian model~\citep{Gal2015} or the variation of the predictions among a number of individual models~\citep{raychaudhuri1995}.

Two major AL categories can be identified in the literature~\citep{ANGLUIN2004175,LangBaum1992,lewis1995sequential}. The methods based on \textit{membership query synthesis} expand an existing data set by creating and labeling new samples that the model is most curious about. 
In contrast, the methods based on~\textit{pool-based sampling} utilize a `bounded' unlabeled data set (also called repository) $\mathcal{U}$, select and  label the most informative samples from $\mathcal{U}$, and include them in the labeled training data set $\mathcal{L}$.
In the present work, pool-based sampling is deemed more suitable as it can prevent creating unrealistic samples~\citep{LangBaum1992}. 
Moreover, identification of the most informative samples follows a query-by-committee (QBC) strategy~\citep{seungetal1992}, in which variance in the outputs of an ensemble of individual models (the committee) is the basis for the next query.
A detailed description of the implemented QBC is provided in section \ref{sec:Metho}.

In summary, the present work aims to answer two questions; first, whether `universal' data-driven predictions of $k_s$ can be approached using a complete statistical-spectral representation of roughness (i.e., with p.d.f. and PS as inputs). We leverage AL to facilitate achieving this goal. The second question is whether and how the information embedded in a data-driven model can provide insight on the contributions of different roughness scales to the added drag. Following this introduction, the roughness generation approach, DNSs, {\color{black} and the} ML methodology are described in section \ref{sec:Metho}. In section \ref{sec:Resu}, first the results and performance of the model are discussed, then the analysis of drag-relevant scales are presented. Section \ref{sec:Conclu} summarizes the main conclusions. 

\section{Methodology}
\label{sec:Metho}
\subsection{Roughness repository}


The (unlabeled) roughness `repository'~$\mathcal{U}$ is constructed by a collection of 4200 artificial irregular rough surfaces.
These surfaces are generated through a mathematical roughness generation method where the PS and p.d.f. of each roughness can be prescribed~\citep{PEREZRAFOLS2019591}. For creation of the present repository, p.d.f. and PS are parameterized, as described shortly, and their parameters are randomly varied within a realistic range to generate a variety of roughness samples while imitating the random nature of roughness formation in practical applications. 

In total, three types of p.d.f. -- namely Gaussian, Weibull, and bimodal -- are used and for each new roughness added to the repository, one type is randomly selected.
The Weibull distribution of random variable $k$ -- here the roughness height -- follows
    \begin{equation}
f_{\text{W}}(k)= K \beta^K k^{(K-1)}e^{-(\beta k)^K},
\end{equation}
where the shape parameter $0.7<K<1.7$ is randomly selected with $\beta=1.0$. In the present notation, $k$ denotes the local roughness height as a function of wall-parallel coordinates $(x,z)$.
The bimodal distribution is obtained combining two Gaussian distributions through~\citep{Peng2000459}:
\begin{equation}
    f_{\text{B}}(k)=f_{\text{G}}(k|0,1)+f_{\text{G}}(k|\mu,\sigma)-f_{\text{G}}(k|0,1)f_{\text{G}}(k|\mu,\sigma),
\end{equation}
where $f_{\text{G}}(x|\mu,\sigma)$ 
is the p.d.f. of the Gaussian distribution with randomized mean $0<\mu<0.5$ and randomized standard deviation $0<\sigma<0.5$.
The p.d.f. variable $k$ is then scaled from 0 to the roughness peak-to-trough height $k_\text{t}=\mathrm{max}(k)-\mathrm{min}(k)$, whose value is randomly determined in the range $0.06<k_\text{t}/H<0.18$, where $H$ is the channel half height.

The PS of the roughness samples in the repository is controlled by 2 randomized parameters, namely the roll-off length $L_r$~\citep{Jacobs_2017} and the power-law decline rate $\theta_{\textrm{PS}}$~\citep{Lyashenko2013}, whose values are selected in the range $0.1<L_r/(\textrm{log}(\lambda_0/\lambda_1))<0.6$ and $-3<\theta_{\textrm{PS}}<-0.1$.
Here, $\lambda_0$ and $\lambda_1$ represent the upper and lower bounds of the PS or the largest and smallest wavelengths forming the roughness topography.
Random perturbations are added to the PS to achieve higher randomness in PS. 
The lower bound of the roughness wavelength is set to $\lambda_1=0.04\mathrm{H}$ to ensure that the finest structures can be discritized by an adequate number of grid points. 
The upper bound of the roughness wavelength $\lambda_0$ is randomly selected in the range $0.5\mathrm{H}<\lambda_0<2\mathrm{H}$.
As will be discussed later, the roughness sample size as well as the simulation domain size should both be adjusted to accommodate this wavelength.

Eventually, 4200 separate pairs of p.d.f. and PS are generated using the described random process, each leading to one rough surface added to the repository $\mathcal{U}$ .
A representation of the parameter space covered by these samples is illustrated in section~\ref{sec:Eva_model}. Moreover, examples of the generated samples can be seen in Appendix \ref{appA}.

\subsection{Direct numerical simulation}
DNS is employed to solve the turbulent flow over selected rough surfaces from the repository in a plane channel driven by constant pressure gradient (CPG). Each DNS leads to determination of the $k_s$ value for the respective roughness sample -- a practice referred to as `labeling' in this paper. 
The DNS is performed with a pseudo-spectral Navier-stokes solver SIMSON~\citep{Chevalier}.
Fourier and Chebyshev series are employed for the discretization in wall-parallel and wall-normal directions, respectively.
Time integration is carried out using third-order Runge-Kutta method for the advective and forcing terms and second order Crank-Nicolson method for the viscous terms.
The roughness representation in the fluid domain is based on the immersed boundary method (IBM) of~\citet{goldstein93}. The code and the IBM are previously validated and used in several publications in the past \citep{PhysRevFluidsPourya,Vanderwel19,yang_stroh_chung_forooghi_2022}.
The solved Navier-Stokes equation writes
       \begin{equation}
       \nabla \cdot \textbf{u}=0~,
        \end{equation}
\begin{equation}
    \frac{\partial \textbf{u}}{\partial t}+\nabla\cdot(\textbf{uu})=-\frac{1}{\rho}\nabla p+\nu\nabla^2\textbf{u}-\frac{1}{\rho}P_x{\mathbf{e_x}}+\textbf{f}_\text{IBM},
\end{equation}
where $\textbf{u}=(u,v,w)^\intercal$ is the velocity vector and $P_x$ is the mean pressure gradient in the flow direction added as a constant and uniform source term to the momentum equation to drive the flow. 
Moreover, $p$, $\mathbf{e_x}$, $\rho$, $\nu$ and $\textbf{f}_\text{IBM}$ denote pressure fluctuation, streamwise unit vector, density, kinematic viscosity and external body force term due to IBM, respectively.
Periodic boundary conditions are applied in the streamwise and spanwise directions. 
The friction Reynolds number is defined as Re$_\tau=u_\tau(\mathrm{H}-k_{\text{md}})/\nu$, where $u_\tau=\sqrt{\tau_w/\rho}$ and $\tau_w=-P_x\cdot (\mathrm{H}-k_{\text{md}})$ are the friction velocity and the wall shear stress, respectively. Here $H$ and $H-k_{\text{md}}$ are half channel height without and with roughness, $k_{\text{md}}$ being the mean (melt-down) roughness height.
In the present work all simulations are performed at Re$_\tau=800$.

Due to the high computational demand of many DNSs, the concept of DNS in minimal channels~\citep{chung_chan_macdonald_hutchins_ooi_2015,MacDonald_2016} is adopted for the considered simulations. 
Recently, ~\citet{yang_stroh_chung_forooghi_2022} showed the applicability of this concept for flow over irregular roughness subject to certain criteria. 
Accordingly, roughness function over a rough surface can be accurately predicted by a comparison of mean velocity profiles in smooth and rough minimal channels if the size of the channels satisfies the following conditions:
 \begin{equation}
 L_z^+\geq \text{max}\left (100,\frac{\Tilde{k}^+}{0.4},\lambda_0^+\right)~,\quad L_x^+\geq \text{max}\left (1000,3L_z^+,\lambda_0^+\right )~.
 \label{MINI2}
 \end{equation}
Here, $L_z$ and $L_x$ are the spanwise and streamwise extent of the minimal channel, respectively, $\lambda_0$ is the largest wavelength in roughness spectrum, and $\Tilde{k}$ is the characteristic physical roughness height. The plus superscript indicates viscous scaling hereafter.
The above condition suggests that, the minimal channel size of each roughness should be determined based on the $\lambda_0$ (which in practice defines the most strict constraint). As described before, $\lambda_0$ is known for each generated roughness sample.  Table~\ref{tab:Sim} summarizes the simulation set up for all DNSs based on the respective $\lambda_0$ value.
Due to the different sizes of the simulation domains, the chosen numbers of grid points differ according to the mesh size, but in all cases $\Delta_{x,z}^+\leq4$.
In wall-normal directions, cosine stretching mesh is adopted for the Chebychev discretization.
The mesh independence is confirmed in a set of additional tests.

For each investigated roughness, $\Delta U^+$ is determined from the offset in the logarithmic velocity profile comparing corresponding rough and smooth DNSs. Notably, when plotting mean velocity profiles, zero-plane displacement $y_0$ is applied in order to achieve parallel velocity profiles in the logarithmic layer, where $y_0$ is determined as the moment centroid of the drag profile on the rough surface following Jackson's method~\citep{jackson_1981} . It is worth noting that in the extensive literature on rough wall-bounded turbulent flows, various definitions of $y_0$ have been proposed, and furthermore, the choice of virtual wall position can affect the predicted rough-wall shear stress $\tau_w$ and thus the resulting $k_s$ value~\citep{chan-braun_garcia-villalba_uhlmann_2011}. Therefore, it is important to recognize this as a possible source of uncertainty and take into account the definitions of $\tau_w$ and $y_0$ when comparing data from different sources.

It is also important to determine if the flow has reached the fully rough regime in each simulation. To this end $\Delta U^+$ is combined with Eqn.~\ref{Eqn_EquivalentSand} to yield a testing value of $k_s^+$. Then following the threshold adopted by~\citet{jouybari_2021} a roughness with $k_s^+\geq50$ is deemed to be in the fully rough regime and all samples not matching this criterion are excluded from the training or testing process. 
The selected threshold of $k_s^+\geq50$ is somewhat lower than the common threshold of $k_s^+\geq70$~\citep{flack2010} and thus may introduce some data points with limited transitionally-rough behavior into the database.
This threshold is, however, deliberately chosen as a trade-off to maximize the number of training data given the limited computational resources. {\color{black}  One should note that an increase in the threshold value of $k_s^+$ while maintaining the same parameter space would be possible by increasing  Re$_\tau$. This would, however, lead to an obvious compromise in the final performance of the model by reducing the number of training data points at a given computational cost.}

Overall, 85 roughness samples are DNS-labeled and eventually included in the labeled data set $\mathcal{L}$ to train the final AL-based model. The procedure for selection of these training samples is explained in detail in the following.
Eight out of the 85 labeled samples are located in the range of $50\leq k_s^+\leq 70$.
We observe that incorporating these samples into the training process improves model performance.
This improvement in the model performance can be attributed both to the incorporation of more informative samples according to AL as well as to the regularization effect of data diversity introduced by including transitionally rough training samples, which makes the model more robust and mitigates over-fitting~\citep{Reed_Nerualsmithing,bishop1995neural}.

\begin{table}
    \centering
    \caption{Simulation setups}
    \begin{tabular}{c|cccccccccc}
         $\lambda_0/\mathrm{H}$& $L_x/\mathrm{H}$ &$L_z/\mathrm{H}$& $L_y/\mathrm{H}$ & $N_x$& $N_z$& $N_y$&$\Delta_x^+$&$\Delta_z^+$&$\Delta_{y,min}^+$&$\Delta_{y,max}^+$  \\
         \hline
         $0.5\geq\lambda_0/\mathrm{H}\geq0.6$&1.8&0.6&2&400&160&451&3.6&3.0&0.02&5.7\\ 
         $0.6>\lambda_0/\mathrm{H}\geq0.7$&2.1&0.7&2&576&144&451&2.9&3.9&0.02&5.7\\
         $0.7>\lambda_0/\mathrm{H}\geq0.8$&2.4&0.8&2&480&192&451&4.0&3.3&0.02&5.7\\
         $0.8>\lambda_0/\mathrm{H}\geq0.9$&2.7&0.9&2&640&256&451&3.4&2.8&0.02&5.7\\
         $0.9>\lambda_0/\mathrm{H}\geq1.0$&3.0&1.0&2&640&256&451&3.8&3.1&0.02&5.7\\
         $1.0>\lambda_0/\mathrm{H}\geq1.1$&3.3&1.1&2&720&288&451&3.7&3.1&0.02&5.7\\
         $1.1>\lambda_0/\mathrm{H}\geq1.2$&3.6&1.2&2&720&288&451&4.0&3.3&0.02&5.7\\
         $1.2>\lambda_0/\mathrm{H}\geq1.3$&3.9&1.3&2&800&320&451&3.9&3.3&0.02&5.7\\
         $1.3>\lambda_0/\mathrm{H}\geq1.4$&4.2&1.4&2&960&384&451&3.5&2.9&0.02&5.7\\
         $1.4>\lambda_0/\mathrm{H}\geq1.5$&4.5&1.5&2&960&384&451&3.8&3.1&0.02&5.7\\
         $1.5>\lambda_0/\mathrm{H}\geq1.6$&4.8&1.6&2&960&384&451&4.0&3.3&0.02&5.7\\
         $1.6>\lambda_0/\mathrm{H}\geq1.7$&5.1&1.7&2&1080&432&451&3.8&3.1&0.02&5.7\\
         $1.7>\lambda_0/\mathrm{H}\geq1.8$&5.4&1.8&2&1200&480&451&3.6&3.0&0.02&5.7\\
         $1.8>\lambda_0/\mathrm{H}\geq1.9$&5.7&1.9&2&1200&480&451&3.8&3.2&0.02&5.7\\
         $1.9>\lambda_0/\mathrm{H}\geq2.0$&6.0&2.0&2&1200&480&451&4.0&3.3&0.02&5.7\\
    \end{tabular}
    
    \label{tab:Sim}
\end{table}
\subsection{Machine learning}
\begin{figure}
    \centering
    \includegraphics[width=.9\textwidth]{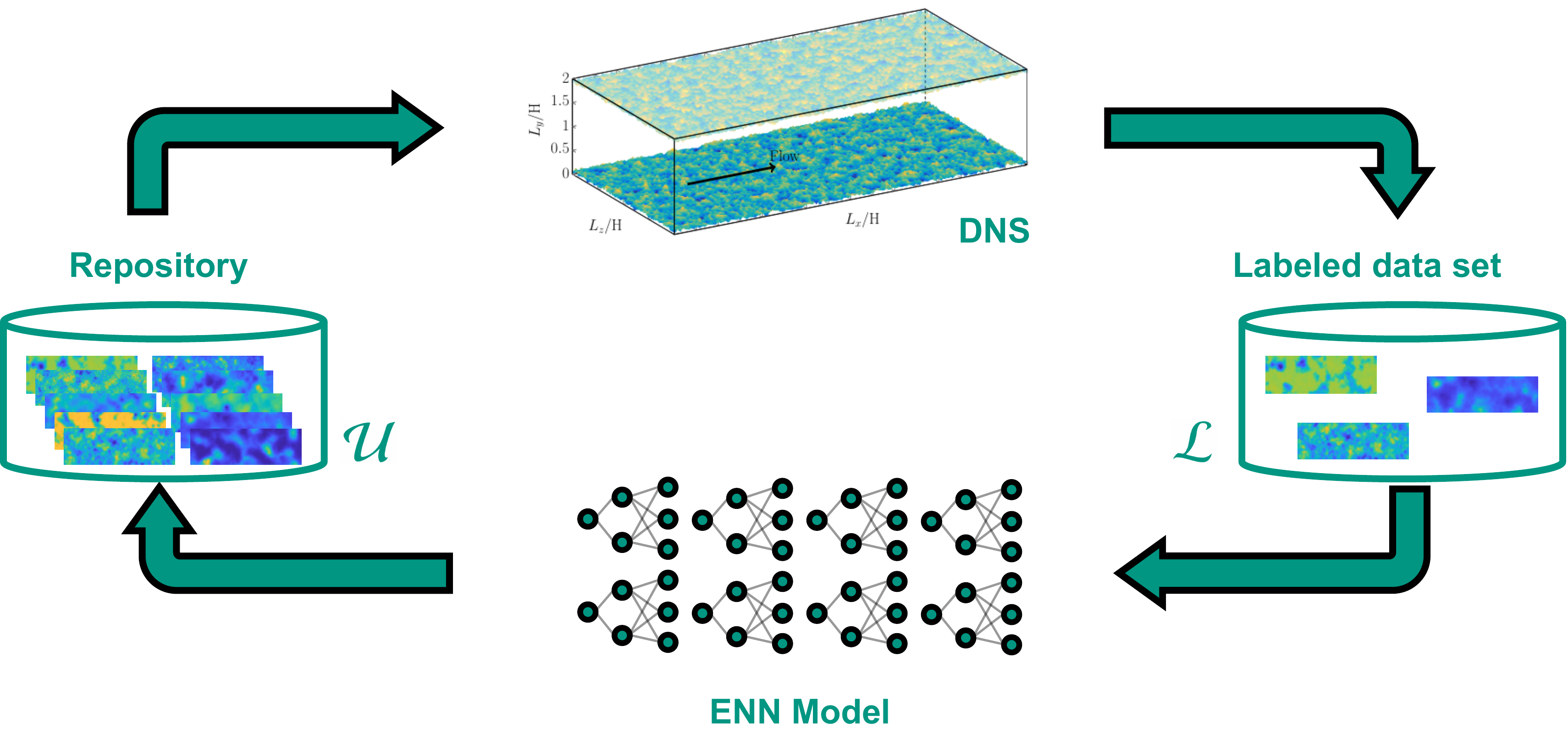}
    \caption{Schematic of the AL-framework.}
    \label{fig:ALLoop}
\end{figure}

The machine learning model in the present work is constructed in a QBC fashion by building an ensemble neural network (ENN) model consisting of 50 independent neural networks (NNs) with identical architecture as the `committee' members.
Similar to the methods proposed by ~\citet{raychaudhuri1995} and \citet{Burbidge2007}, the prediction uncertainty of the ENN model is defined as the variance of the predictions among the members, $\sigma_{k_r}$.

The workflow of the AL framework is sketched in the figure~\ref{fig:ALLoop}.
Two collections of roughness samples are included in the framework. These are the (unlabeled) repository $\mathcal{U}$ and the (labeled) training data set $\mathcal{L}$.
As a starting point in the AL framework, 30 samples are randomly selected from the repository, labeled (i.e. their $k_s$ is calculated) through DNS, and used to train a first ENN model, which is referred to as the `base model'. 
This preliminary `base model' is subsequently improved throughout multiple AL iterations.
In each AL iteration, approximately 20 new roughness samples from $\mathcal{U}$ are DNS-labeled and added to $\mathcal{L}$ for training of ENN. These are the samples in $\mathcal{U}$ with the highest prediction variances according to the most recent ENN. This QBC strategy leads to an effective exploration of the repository and adding the new data at the most uncertain regions of the parameter space. 

\begin{figure}
    \centering
    \includegraphics[width=\linewidth]{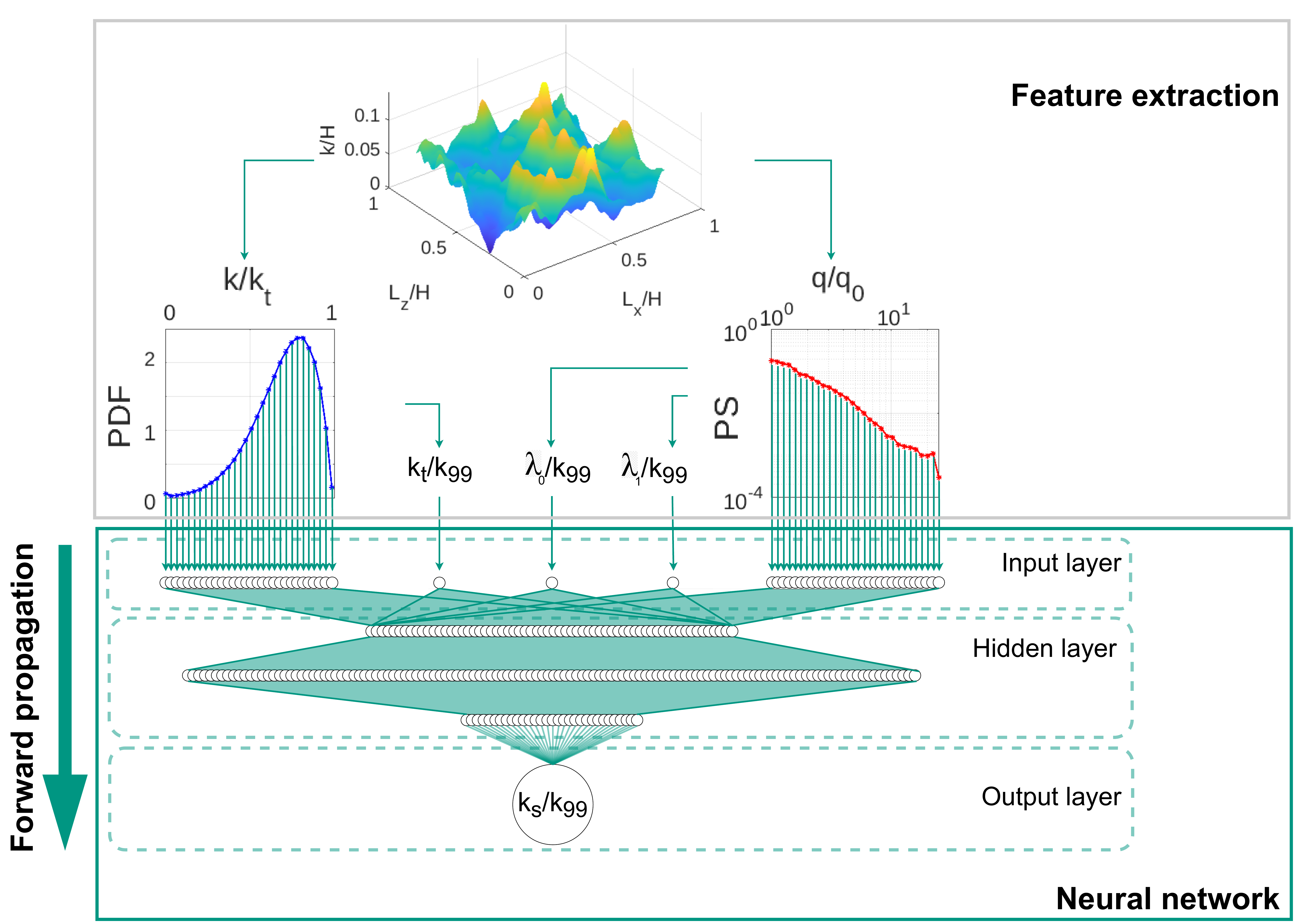}
    \caption{Schematic of a single neural network in ENN.}
    \label{fig:DNN}
\end{figure}
The function of the ENN model is to regress the (dimensionless) equivalent sand-grain roughness $k_r=k_s/k_{99}$, and to calculate the variance of the predictions $\sigma_{k_r}$ as a basis for QBC ($k_{99}$ is the 99\% confidence interval of the roughness p.d.f., which is used as the representative physical scale of roughness height in this paper). The ENN is composed of multiple NNs with similar structures that is shown in figure~\ref{fig:DNN}.
The input vector $\bm{I}$ of the NN contains the discretized roughness p.d.f. and PS along with 3 additional characteristic features of the rough surface, i.e. $k_\text{t}/k_{99}$ and the normalized largest and smallest roughness wavelength $\lambda_0^*=\lambda_0/k_{99}$ and $\lambda_1^*=\lambda_1/k_{99}$, respectively.
The input elements in $\bm{I}$ that represent the roughness p.d.f. and PS are obtained by equidistantly discretizing the roughness p.d.f. and PS each into 30 values within the height range $0<k<k_\text{t}$ and the wavenumber range $2\pi/\lambda_1>2\pi/\lambda>2\pi/\lambda_0$.
Each NN in the ensemble is constructed with one input layer with 63 (3+30+30) input elements, three hidden layers with 64, 128 and 32 non-linear neurons with rectified linear units (ReLUs) activation ($\text{max}\{0,x\}$), and one linear neuron in the output layer.
The optimal number of neurons at each layer is determined through a grid search of a range of numbers that achieves the lowest model prediction error on $\mathcal{T}_{\text{inter}}$.
L2-regularization is applied to the loss function. Adaptive momentum estimation (Adam) is employed to train the model.
The final prediction of the ENN is defined as the mean prediction over the 50 NNs, namely $\mu_{k_r}=\sum_{i=1}^{50}\hat{k}_{r,i}/50$, where $\hat{k}_r$ represents the prediction of a single NN, the index $i$ indicates the index of the NN.
The prediction variance is calculated as  $\sigma_{k_r}=\sqrt{\sum_{i=1}^{50}(\hat{k}_{r,i}-\mu_{k_r})^2/50}$.
It is worth noting that each NN in the ENN model is individually trained based on 90\% of the randomly selected samples in the labeled data set $\mathcal{L}$ while the rest of the samples are used for validation.
The initial weights of the neurons in each NN are randomly assigned at the beginning of the training process.
In such a way, the diversity among the QBC members is ensured, which is an important factor in determining the generalization of the ENN model~\citep{MelvilleMooney2003}.
{\color{black} It is important to note that the current ensemble members used in the model are deterministic neural networks, and the uncertainty of the training data from DNS is assumed to be minimal. However, when considering experimental training data, where (aleatoric) uncertainties arise from possible measurement errors, the performance of the current ENN approach may be compromised due to its limited capability in handling such uncertainties. In these scenarios, the utilization of probabilistic models -- such as Bayesian neural networks -- may be more suitable as they allow for the explicit incorporation of measurement uncertainties.}
\subsection{Testing data sets}
In the present work, three distinct testing data sets are introduced to evaluate the model performance and its universality.
The difference among the data sets lies in the nature and origin of samples they contain. 
The first data set $\mathcal{T}_{\text{inter}}$ is composed of 20 samples randomly chosen from $\mathcal{U}$ that have been never seen by the model during the training process.

Despite the fact that the employed roughness generation method can generate irregular, multi-scale surfaces resembling realistic roughness, we separately test the model for additional rough surfaces extracted from scanning of naturally-occurring roughness, which form the second testing data set $\mathcal{T}_{\text{ext,1}}$. There are five samples in this `external' data set. These include roughness generated by ice accretion~\citep{Juan2019}, deposit in internal combustion engine~\citep{FOROOGHI201883}, and a grit-blasted surface~\citep{Thakkar2017}.
In addition to that, we test the model against a second external data set $\mathcal{T}_{\text{ext,2}}$, which contains irregular roughness samples from the database provided by~\citet{jouybari_2021}. In this data set, many roughness samples are generated by placing ellipsoidal elements of different sizes and orientations on a smooth wall, making them rather distinct from the type of roughness used to train the model. We separate this testing data set from the other two as it contains a specific type of artificial roughness. 

\section{Results}
\label{sec:Resu}
\subsection{Assessment of the active learning framework}
\label{sec:Eva_model}

\begin{figure}
    \centering
    \begin{tikzpicture}
    \begin{groupplot}[group style={group size= 3 by 1}]
        \nextgroupplot[
        			xlabel=$2\pi k_{99}/\lambda$,
		ylabel=PS,
		xmin=.01,xmax=10,
		ymin=0.0000001,ymax=1,
		clip=true,
		set layers,
		xmode=log,
		ymode=log,
		clip mode=individual,
		height=.46\textwidth,
		width=.46\textwidth,
        tick label style={font=\footnotesize},
                    legend style={font=\tiny,anchor=south west},
                        legend pos=south west,
	]
	  \addplot [
            gray,thick
            ]
            coordinates{
            (100,2)			
			(100,1)
			};
			\label{repository}
  \addplot [
            kit-green100,thick
            ]
            coordinates{
            (100,2)			
			(100,1)
			};
			\label{labeled_0}
			              \addplot [
            red,thick
            ]
            coordinates{
            (100,2)			
			(100,1)
			};
			\label{labeled_1}
              \addplot [
            blue,thick
            ]
            coordinates{
            (100,2)			
			(100,1)
			};
			\label{labeled_2}
              \addplot [
            black,thick
            ]
            coordinates{
            (100,2)			
			(100,1)
			};
			\label{labeled_furtherround}
\addplot [thick, color=blue, on layer=axis background]
graphics[xmin=.01,ymin=0.0000001,xmax=10,ymax=1]{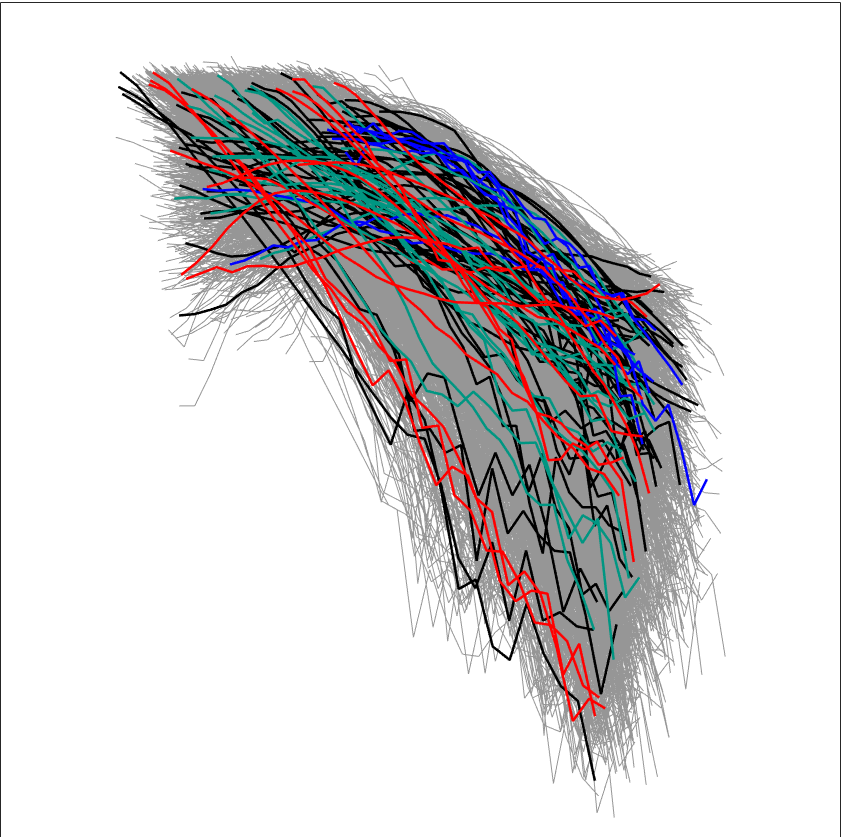};

\node[below right] at (current axis.north west)
{{(a)}};
 \coordinate (top) at (rel axis cs:0,1);
 \nextgroupplot[width=.15\textwidth,
            height=0.2\textwidth,
                  axis line style={draw=none},
      tick style={draw=none},
      xticklabels={,,},
      yticklabels={,,}
            ]
        \nextgroupplot[
            xlabel=$k/k_t$,
		ylabel=p.d.f.,
		xmin=0,xmax=1,
		ymin=0,ymax=25,
		clip=true,
		set layers,
		clip mode=individual,
		xtick={0,0.2,0.4,0.6,0.8,1},
		        ylabel shift = -5 pt,
		height=.46\linewidth,
		width=.46\linewidth,
        tick label style={font=\footnotesize},
                    legend style={font=\footnotesize,anchor=north east},
                        legend pos=north east,
	]
	\centering

\addplot [thick, color=blue, on layer=axis background]
graphics[xmin=0,ymin=0,xmax=1,ymax=25]{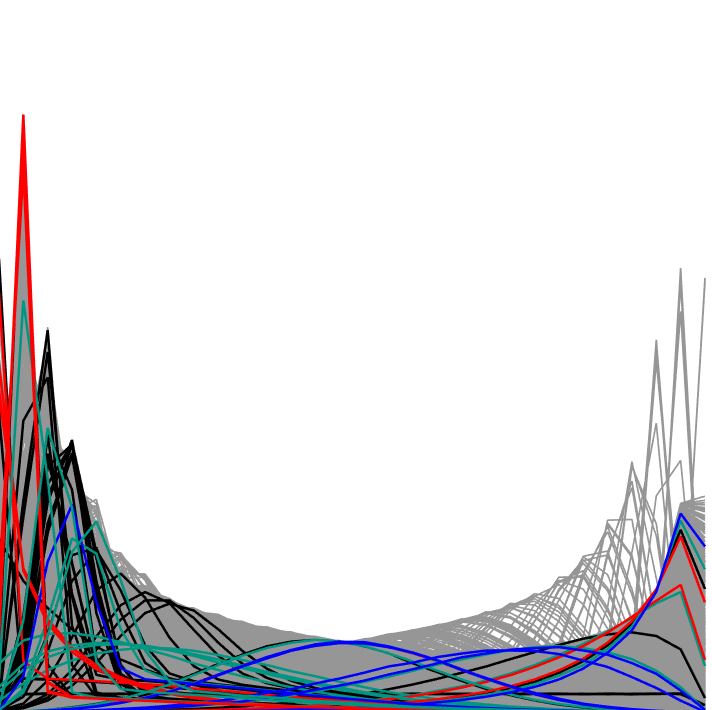};
\node[below right] at (current axis.north west)
{{(b)}};
            \coordinate (bot) at (rel axis cs:1,0);
    \end{groupplot}
    \path (top-|current bounding box.west)-- 
          node[anchor=south,rotate=90]{}
          (bot-|current bounding box.west);
\path (top|-current bounding box.north)--
      coordinate(legendpos)
      (bot|-current bounding box.north);
\matrix[
    matrix of nodes,
    anchor=south,
    draw,
    inner sep=0.2em,
    draw
  ]at([yshift=1ex]legendpos)
  {
    \ref{repository}& \footnotesize Roughness repository&[2pt]
    \ref{labeled_0}& \footnotesize Initial database&[2pt]
    \ref{labeled_2}& \footnotesize EQ selected,first round\\
    \ref{labeled_1}& \footnotesize AL selected, first round&[2pt]
    \ref{labeled_furtherround}& \footnotesize AL selected, further rounds\\};
\end{tikzpicture}
    \caption{PS (a) and p.d.f. (b) of 4200 roughness samples in the roughness repository (gray). The samples selected for training are distinguished with different colors. While AL model tends to explore the PS and p.d.f. domain, the EQ model contains samples that are placed closely to the known initial database.}
    \label{fig:PDFPS}
\end{figure}
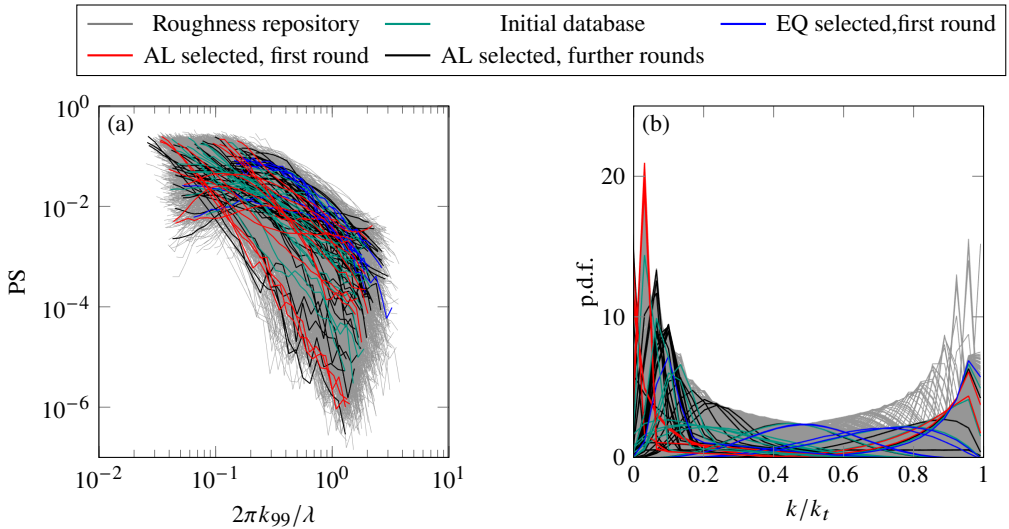

\begin{figure}
    \centering
    \input{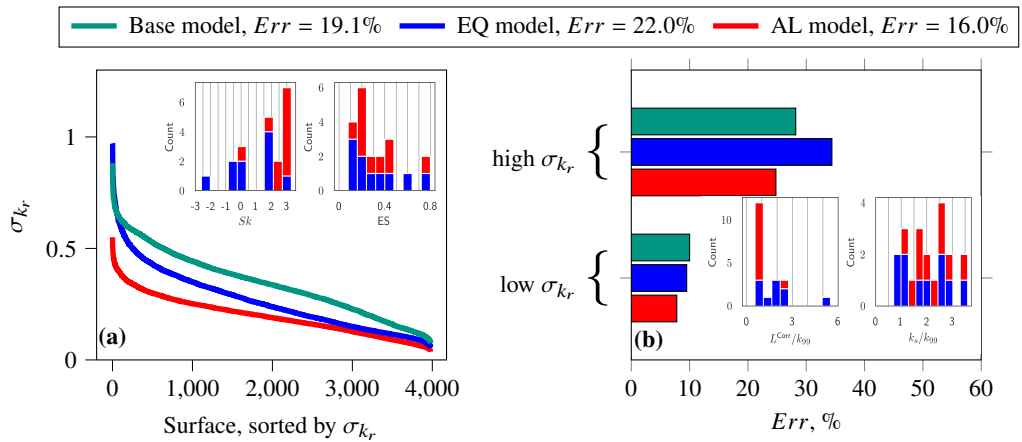}
    \caption{(a) Prediction variance $\sigma_{k_r}$ obtained by three different models for all the samples in repository $\mathcal{U}$. (b) The average error obtained by the three models for 10 high-variance samples and 10 low-variance samples in $\mathcal{T}_\text{inter}$ (sorted based on the variance of the base model). The total averaged errors are displayed in the legend. Insets: The distribution of the statistical parameters as well as the corresponding $k_r$ of the new samples with AL and EQ sampling strategies with identical color code.}
    \label{fig:ALVSEQ}
\end{figure}

In this section, we explore if the AL framework enhances the training behavior of the model. To do so, we compare a model trained with AL-selected data points to one trained with an arbitrary selection of data points. To avoid computational cost of running many eventually unused DNSs, the comparison is made for only one AL iteration. Figure~\ref{fig:PDFPS} shows all p.d.f. and PS pairs contained in the repository $\mathcal{U}$ (gray) and those randomly selected for the initial `base model' (green), as well as those selected for further training (other colors).
The wide range of available roughness can be understood from the area covered by gray curves.
As explained before, once the base model is trained using the initial randomly selected data set, it is used to determine which samples from the repository $\mathcal{U}$ should be selected for the next round of training. In figure~\ref{fig:ALVSEQ}(a) the green line shows the prediction variance $\sigma_{k_r}$ of all roughness samples in the repository based on the base model. Here the abscissa is the sample number sorted from high to low $\sigma_{k_r}$ values. According to the AL framework, the samples selected for the next round are the ones with the largest $\sigma_{k_r}$. These are shown with red color in figure~\ref{fig:PDFPS}. For comparison, a second sampling strategy (denoted as EQ) is employed in which the same number of samples as in AL are selected, but they are equidistantly distributed along the abscissa of figure~\ref{fig:ALVSEQ}(a). These samples are shown with blue color in figure~\ref{fig:PDFPS}.
It is clearly observed in figure~\ref{fig:PDFPS} that the AL model explores surfaces that are least similar to those in the initial data set (green) and tend to cover the entire repository with a higher weight given to the marginal cases.
Furthermore, the parameter distribution as well as the corresponding $k_r$ values of the selected roughness by means of AL and EQ is compared in the insets of figure~\ref{fig:ALVSEQ}.
{\color{black}It can be seen, that both the AL and EQ models generally prioritize selecting samples within the waviness regime i.e. ES$<0.35$~\citep{napoli_armenio_demarchis_2008}. This preference may arise from the fact that the resulting drag in the waviness regime (ES$<0.35$) is sensitive to changes in ES~\citep{Schultz2009}. Conversely, beyond this regime (ES $>$ 0.35), the resulting $\Delta U^+$ saturates in relation to increasing ES, making these samples less interesting for both labeling strategies.}
{\color{black}On the other hand}, the AL model {\color{black}particularly} tends to sample the roughness with positive skewness and low correlation length. This can {\color{black}similarly} be a result of the roughness effect being highly sensitive to the variations in roughness statistics within these range of parameters, which is in-line with previous findings~\citep{Busse2023,Schultz2009}.

Subsequently, two separate models are trained based on the AL and EQ strategies. These models are applied separately to determine the variance of prediction for roughness in the repository, and the results are depicted in figure~\ref{fig:ALVSEQ}(a) using red and blue lines. It is evident from the results that both the AL and EQ models generally reduce the prediction variance. 
However, a more substantial decline in the values of $\sigma_{k_r}$ is achieved by the AL model. This is the expected behavior as AL is designed to reduce the prediction uncertainty by targeting regions of the parameter space where the uncertainty is the largest. Interestingly, some increase in $\sigma_{k_r}$ of the EQ model can be observed for a number of samples with very high $\sigma_{k_r}$, which can be a sign that the performance of EQ model in the `difficult' tasks deteriorates as it is not trained well for those tasks due to ineffective selection of its training data.
Moreover, the prediction errors (calculated based on correct $k_s$ values of testing data set $\mathcal{T}_{\text{inter}}$ obtained by DNS) are illustrated in figure~\ref{fig:ALVSEQ}(b).
The averaged prediction errors, $Err$, achieved by the base model, the AL model, and the EQ model for the entire $\mathcal{T}_{\text{inter}}$ are {\color{black}19.1\%, 16.0\%, and 22.0\%}, respectively. 
While AL model yields a meaningful reduction in $Err$, the overall performance of the EQ model deteriorates, possibly due to the over-fitting, which in our case refers to the condition where the model is trained to fit a limited number of relatively similar data points so precisely that its ability to extrapolate on dissimilar testing data is degraded~\citep{Hastie2009}. 
To better analyse this observation, the testing data set $\mathcal{T}_\text{inter}$ is evenly split into 2 subsets according to their $\sigma_{k_r}$, namely the high- and low-variance subsets. 
The $Err$ values for both high and low-variance subsets are illustrated in the figure.
It is clear that, while the EQ strategy improves the model performance for the already low-variance test data, its error increases for high-variance test data, which can be taken as an indication of over-fitting as described above.
The AL sampling strategy, in contrast, seems to protect the model from over-fitting -- especially in the circumstance of a small training data set -- hence the error is reduced for both high- and low-certainty test data as a result of effective selection of training data.

\subsection{Performance of the final model}

\begin{figure}
    \centering
    \includegraphics[width=.85\linewidth]{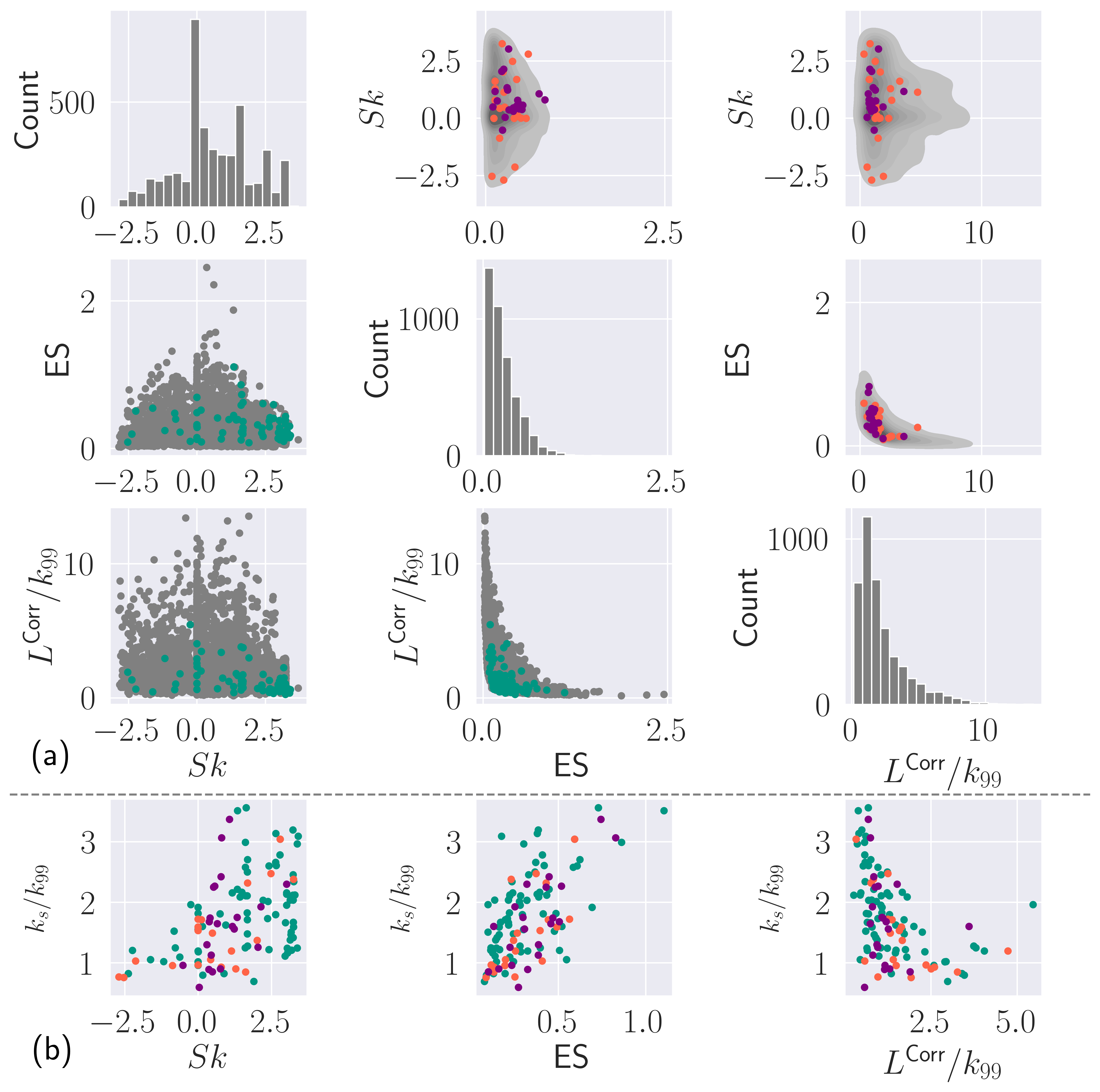}
    \caption{(a) Pair plots of roughness statistics. Lower left: The distribution of the samples in $\mathcal{U}$ (gray) and $\mathcal{L}$ (green). Diagonal: Histogram of single roughness statistics in  $\mathcal{U}$. Upper right: Joint probability distribution of statistics overlaid by test data in $\mathcal{T}_{\text{inter}}$ (orange) and $\mathcal{T}_{\text{ext,1\&2}}$ (purple). (b) Values of $k_r=k_s/k_{99}$ obtained from DNS (ground truth) as a function of the selected statistics. Color code is same as in (a)}
    \label{fig:Pairplot}
\end{figure}
\begin{figure}
    \centering
    \includegraphics[width=.7\linewidth]{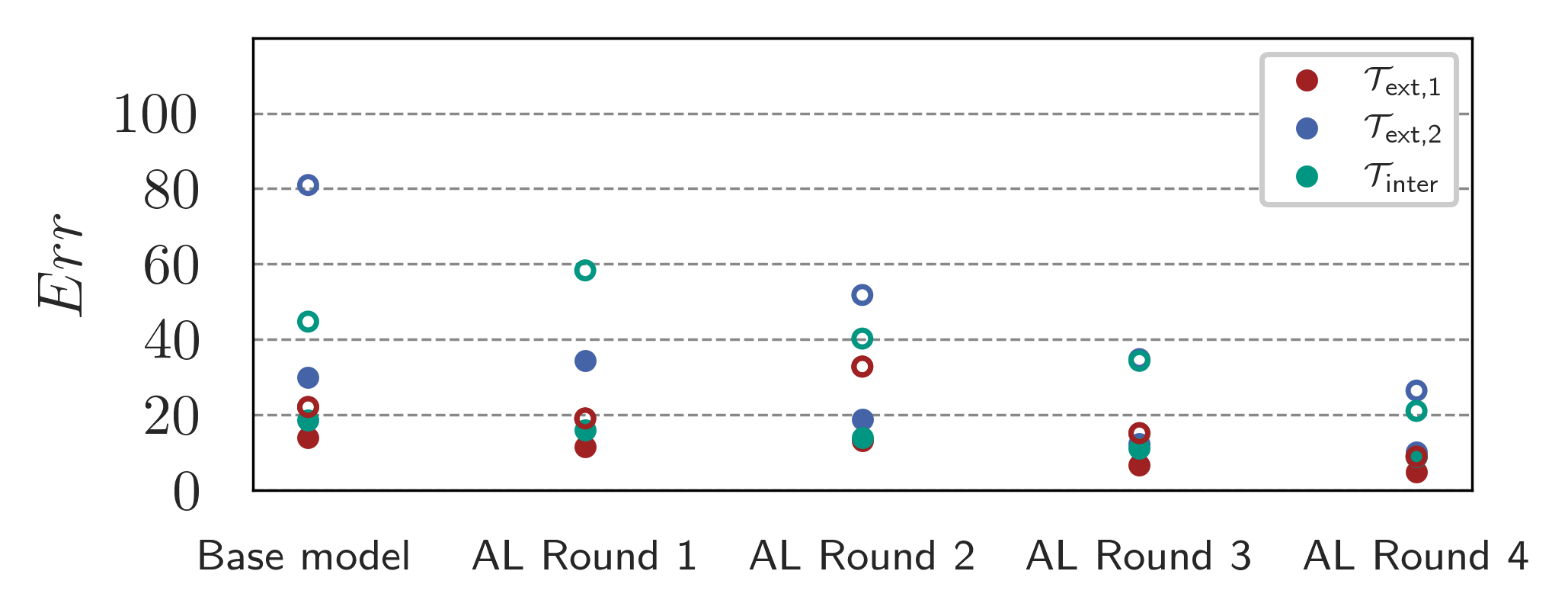}
    \caption{The arithmetically averaged $Err$ [\%] as well as maximum $Err$ of the model after different training rounds on each of the testing data sets $\mathcal{T}_\text{inter}$, $\mathcal{T}_\text{ext,1}$, $\mathcal{T}_\text{ext,2}$. The mean $Err$ is represented with closed circle while the maximum $Err$ is displayed with open circle with corresponding color. The maximum $Err$ for $\mathcal{T}_{\text{ext,2}}$ at AL Round1 is out of the plot range.}
    \label{fig:ErrEvoTesting}
\end{figure}
Having demonstrated the advantage of AL over random sampling, three additional AL iterations are carried out.
The distribution of the PS and p.d.f. of the selected roughness from the second to the fourth AL iterations are displayed in figure~\ref{fig:PDFPS} with black lines. 
A number of roughness maps from each AL round are also displayed in Appendix~\ref{appA}.

The total number of data points for training of the model after four iterations adds up to 85; these are the data that form $\mathcal{L}$. The scatter plots of some widely investigated roughness parameters in $\mathcal{L}$ as well as in the unlabeled repository $\mathcal{U}$ are displayed in the lower left part of figure~\ref{fig:Pairplot}(a).
In the figure, $k(x,z)$ is the elevation map of the roughness, $Sk=1/(Sk_{\text{rms}}^3)\int_S(k-k_{\text{md}})^3\mathrm{d}S~$ represents the skewness, where $S$ is the wall-projected surface area, and $k_{\text{md}}=1/S\int_Sk\mathrm{d}S~$ is the melt-down height of the roughness. The effective slope is defined as ES $=1/S\int_S|(\partial k)/(\partial x)|\mathrm{d}S~$. $L^{\text{Corr}}$ is the correlation length representing the horizontal separation at which the roughness height auto-correlation function drops under 0.2.
An inverse correlation can be observed between $L^{\text{Corr}}$ and ES, which is expected as roughness with larger dominant wavelength tend to have lower mean slope.
The distribution of other statistics in $\mathcal{U}$ appears to be reasonably random.

For the sake of comparison, the test data are additionally represented in the upper right part of the plots with orange (for $\mathcal{T}_{\text{inter}}$) and purple (for $\mathcal{T}_{\text{ext, 1\&2}}$) symbols. 
It is worth noting that only the roughness samples that locate in the fully rough regime at the currently investigated Re$_\tau$ are included in $\mathcal{L}$ and shown in the figure. 
Figure~\ref{fig:Pairplot}(b) shows the values of $k_r$ (from DNS) against the three roughness statistics for all labeled data in the training and testing data sets.
As can be clearly observed in the figure, while equivalent sand-grain roughness shows some general correlation with each of these statistics (increasing with $Sk$ and ES, decreasing with $L^{\text{Corr}}$), the collapse of data is far from perfect. Clearly, no roughness statistics can entirely capture the effect of an irregular multi-scale roughness topography on drag, which is essentially a motivation behind seeking a NN-based model to find the functional relation between $k_s$ and a higher order representation of roughness (here p.d.f. and PS). 

Eventually, the final model is trained on the entire labeled data set $\mathcal{L}$. The mean and maximum error values achieved by this model on all three testing data sets, as well as those errors after each training round, are displayed separately in figure~\ref{fig:ErrEvoTesting}. The figure shows a generally decreasing trend in both mean and maximum error as the model is progressively trained for more AL rounds, despite some exceptions to the general trend in the first two rounds when the number of data points is low. It is notable that the AL model is particularly successful in bringing down the maximum error, and hence, can be considered reliable over a wide range of scenarios.

One should mention that the model performs consistently well for three different testing data sets with different natures. While the data set $\mathcal{T}_{\text{inter}}$ covers an extensive parameter space -- hence containing more extreme cases -- it is generated employing the same method as the training data. Therefore, to avoid a biased evaluation of the model, two `external' testing data sets from literature are also included. The data set $\mathcal{T}_{\text{ext,2}}$ is believed to be particularly challenging for the model, since it is formed by roughness generated artificially using discrete elements \citep{jouybari_2021}, which is fundamentally different from the target roughness of this study. Nevertheless, the final model yields very similar errors for all data sets; what can be taken as an indication of its generalizability. The averaged errors of the final model within the data sets $\mathcal{T}_{\text{inter}}$, $\mathcal{T}_{\text{ext,1}}$, and $\mathcal{T}_{\text{ext,2}}$ are approximately {\color{black}9.3\%}, 5.2\%, and 10.2\%, respectively.

{\color{black}It is crucial to acknowledge that the present model is developed under the assumption of statistical surface homogeneity. However, when reaching beyond this assumption, the presence of surface heterogeneity introduces additional complexity to the problem that cannot be adequately represented by the current training samples. As a consequence, the effect of heterogeneous roughness structures~\citep{Hinze1967,stroh_schaefer_frohnapfel_forooghi_2020} cannot be adequately accounted for by the current model.}
\subsection{Data-driven exploration of drag-relevant roughness scales}
\label{sec:LRP}
The fact that naturally-occurring roughness usually has a multi-scale nature with continuous spectrum is well established~\citep{SAYLES1978}. How spectral content of roughness affects skin-friction drag and whether a certain range of length-scales dominates it are, however, questions receiving attention more recently~\citep{anderson_meneveau_2011,Mejia2010,BARROS20181,medjnoun2021}. 
In this sense,~\citet{BUSSE2015129} applied low-pass Fourier filtering to a realistic roughness and observed no significant effect on skin-friction drag when the filtered wavelengths were lower than a certain threshold. On the other hand,~\citet{BARROS20181}, used high-pass filtering and suggested that very large length scales may not significantly contribute to drag. ~\citet{Alvesportela21} examined three filtered surfaces, each maintaining one-third of the original spectral content associated with large, intermediate, or small scales. In all cases, the filtered scales were shown to include `drag-relevant' information. While both lower and higher limits of drag-relevant scales (if they exist) can be matter of discussion, the present study mainly focuses on the latter. Possibly related to that question,~\citet{Schultz2009} documented the equivalent sand-grain size of pyramid-like roughness with wavelengths higher (hence  lower effective slopes) than a certain value not to scale in the same way as those with smaller wavelengths. These authors coined the term `wavy' for the high-wavelength roughness behavior. Later on,~\citet{yuan14} revealed that the wavy regime may emerge at a different threshold (in terms of effective slope) in a multi-scale roughness compared to the single-scale pyramid-like roughness. Recently,~\citet{yang_stroh_chung_forooghi_2022} showed that the spectral coherence of roughness topography and time-averaged drag force on a rough wall drops at large streamwise wavelengths, which, in-line with the finding of~\citet{BARROS20181}, suggests decreasing drag-relevance of large scales. 

In the present work, we are particularly interested to explore the possibility of extracting the drag-relevant scales from the knowledge embedded in the data-driven model.
In doing so, we employ the layer-wise relevance propagation (LRP) technique~\citep{Bach2015}, which has previously proven successful in other contexts as a way to interpret decisions of NN models~\citep{DBLP:journals/corr/ArrasHMMS16a,SamekBMBM15}.
LRP is an instance-based technique, which can be used to quantify the contribution of each input feature (here points in discretized p.d.f. and PS) to the output of the model (here $k_r=k_s/k_{99}$) for a single test case (here a roughness sample). 
According to this technique, the contribution score (or relevance) of neuron $j$ at each layer of the deep neural network can be expressed as
\begin{equation}
    R_j=\sum_l \left (\frac{a_jw_{jl}}{\sum_{j}a_jw_{jl}} \right )R_l,
    \label{eqn_LRP}
\end{equation}
where $R_l$ is the contribution score of neuron $l$ in the subsequent layer. In equation \ref{eqn_LRP}, $w$ and $a$ are the weight and activation of the neuron that are obtained when the model is used to predict one instance (here the $k_r$ for the roughness sample of interest). Note that in our NN, the last layer corresponds to the predicted output and the first layer to the input roughness information. For better interpretability we assign the value of one to the contribution score (or relevance) of the output neuron. As a result, the sum of contribution scores of all inputs must be one.
Note that the contribution scores shown in this section are averaged over the 50 NN members.

\begin{figure}
    \centering
    \includegraphics[width=.8\linewidth]{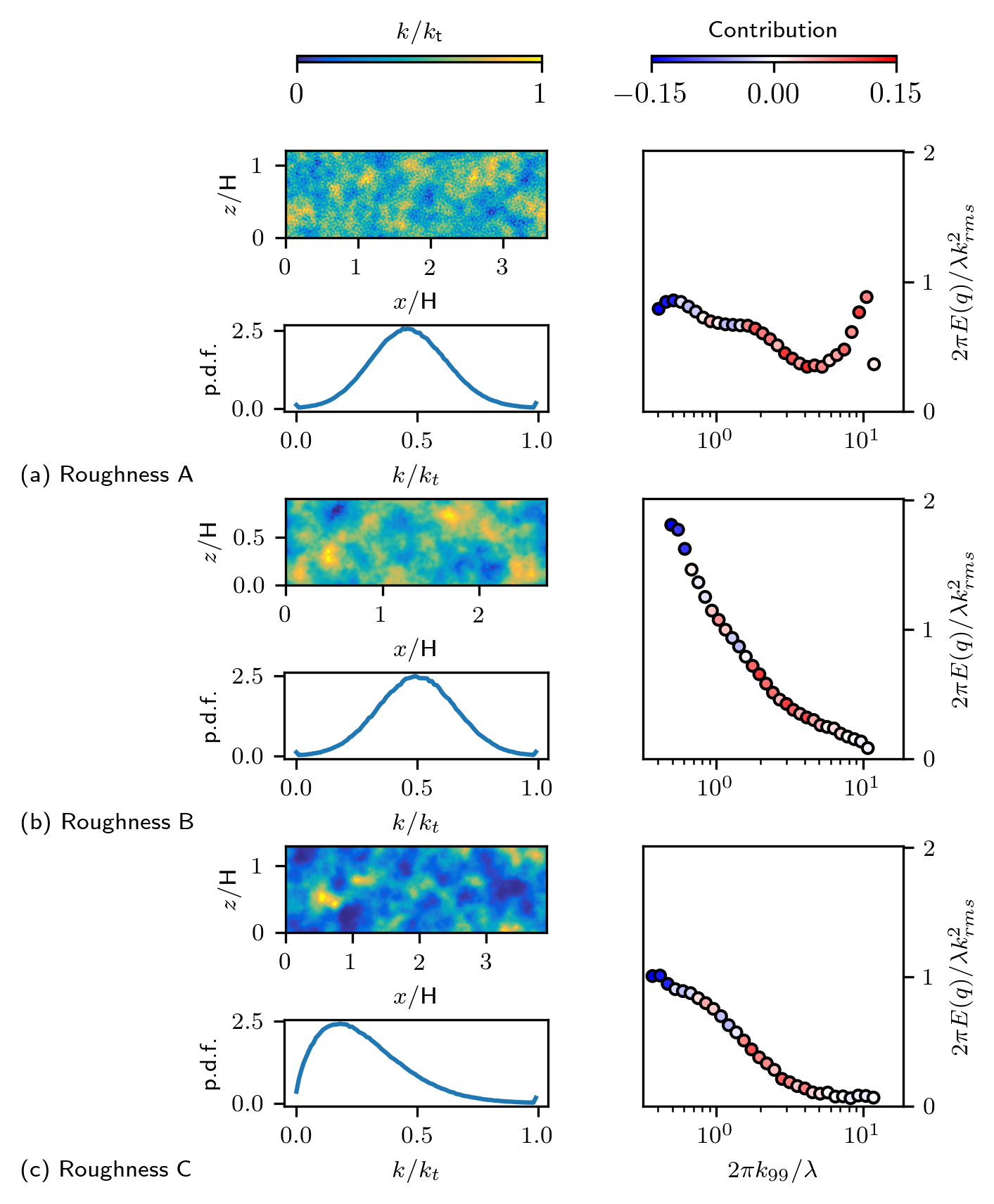}
    \caption{Height maps, p.d.f.s and discretized color-coded pre-multiplied roughness height PS of three exemplary samples A (a), B (b), and C (c). The PS are colored by the LRP contribution scores.}
    \label{fig:LRPPSPDF}
\end{figure}

In order to extract drag-relevant scales we consider the following idea. 
A wavelength that does not affect $k_s$ (which is a measure of added drag) still contributes to an increasing variance of the roughness height, and hence $k_{99}$. Therefore, the related output of the NN, which is the ratio $k_s/k_{99}$, is decreased. 
An input that decreases the output shows a negative LRP contribution score. With that in mind, figure~\ref{fig:LRPPSPDF} shows three exemplary roughness samples (named A, B and C) and their discretized PS. Each discrete wave-number in a PS is an input to the model, thus has a contribution score, which is indicated using the specified colour code. The spectra are shown in pre-multiplied form and the p.d.f. of each roughness is also displayed. Samples with both Gaussian and non-Gaussian p.d.f.s are included.
It is observed in figure~\ref{fig:LRPPSPDF} that the small wave-numbers (i.e. large wavelengths) generally have more negative contribution scores, which is in accordance to the suggestion of~\citet{BARROS20181}. Indeed the most negative contributions consistently belong to the largest wavelengths for all samples. 
On the other hand, smaller wavelengths generally show larger contribution scores, but the trend is not monotonic. This might indicate that drag-relevant scales reside within a certain range of the spectral content.

\begin{figure}
    \centering
    \includegraphics[width=\linewidth]{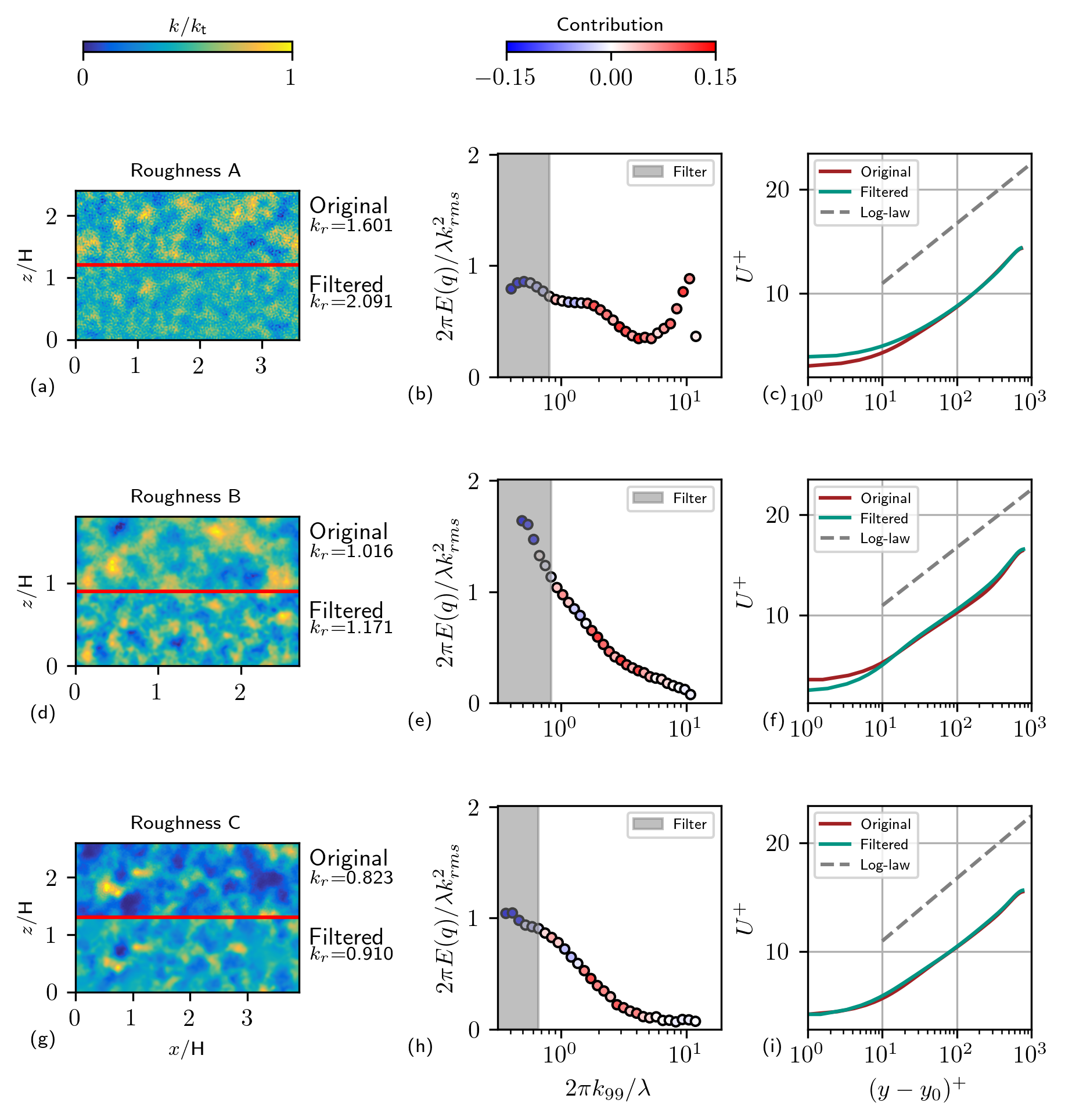}
    \caption{The original and high-pass filtered roughness (a, d, g), the pre-multiplied roughness height PS with the filtered scales indicated by gray shade (b, e, h), and inner-scaled mean velocity profiles out of DNS on the original and filtered roughness (c, f, i). Note that the DNS are carried out in minial channels.}
    \label{fig:LRP_filter}
\end{figure}


\begin{table}
    \centering
    \begin{tabular}{l c c c c c c}
         &$k_\text{md}$/H&$k_{99}$/H&$Sk$&ES&$L^\text{Corr}$/H\\
         \hline
         Roughness A original& 0.043&0.076&0.149&0.520&0.149\\
Roughness A filtered&0.052&0.065&0.115&0.519&0.092\\
         
    \hline
    Roughness B original&0.041&0.069&0&0.146&0.198\\
    Roughness B filtered&0.024&0.041&0.170&0.128&0.102\\
    \hline
    Roughness C original&0.026&0.078&0.885&0.129&0.234\\
    Roughness C filtered&0.034&0.056&0.307&0.122&0.126\\
    \end{tabular}
    \caption{Statistical properties of selected surfaces A, B and C}
    \label{tab:statistics_filtered}
\end{table}
In order to examine whether or not negative LRP contribution score indeed indicates drag-irrelevance, we apply high-pass filtering to the samples in figure~\ref{fig:LRPPSPDF}, and examine the resulting roughness using DNS under the same conditions as for the original roughness. The position of filter is chosen to be the largest wavelength with non-positive contribution score (a 3-point moving average is applied to smoothen the LRP scores beforehand). Figure~\ref{fig:LRP_filter} shows the height map of original versus filtered samples, the spectra with filter positions, and the inner-scaled mean velocity profiles before and after filtering for samples A, B, and C. Some statistical properties of all original and  filtered samples are also displayed in table \ref{tab:statistics_filtered}. It is clear from figure~\ref{fig:LRP_filter} that the velocity profiles of original and filtered samples collapse very well in the logarithmic region and beyond, which obviously leads to similar values of roughness function and drag coefficient. This observation lends support to the hypothesis that the large roughness scales beyond a threshold do not have a meaningful contribution to the added drag, and that LRP analysis can be a data-driven route to identifying those scales {\color{black}\textit{a priori}}.
{\color{black} One obvious application of this finding can be in selection of sampling size for the investigations of roughness effect. In practice, it is not always possible to obtain roughness samples, that are large enough to encompass the full spectrum of scales. However, once the range of drag-relevant scales is completely covered by a roughness sample, a miscalculation due to a limited sample size can be avoided.} 


Interestingly, in all samples shown in figure~\ref{fig:LRP_filter}, the filtered scales have a significant contribution to the roughness height variance based on the pre-multiplied roughness spectra. 
This is also reflected in the significant decrease in roughness height $k_{99}$ and $L^{\text{Corr}}$ in table \ref{tab:statistics_filtered}, as anticipated.
Additionally, based on the three observed cases, the reductions in the ES values are found proportional to the filtered fraction of the PS.
Other statistical parameters also undergo changes due to filtering, while obviously none of these changes are relevant in determining the drag. 
It is worth noting that additional to the roughness height $k_{99}$ and ES, other drag-determining quantities, such as $Sk$, undergo a general reduction for roughness C.
According to some existing empirical correlations (e.g. the correlations proposed by~\citet{chan2015,10.1115/1.4037280,Flack2020}) the simultaneous reduction in these statistics should lead to a lower $k_s$. This is however not the case in reality based on the DNS results which can be reminiscent of the suggestion by~\citet{BARROS20181} that a high-pass filtering is necessary if predictive correlations are to capture the correct trend between $k_s$ and the roughness statistics.This also provides an indication for the hypothesis that, while statistical parameters can correlate the equivalent sand grain size of irregular roughness to some degree, only a combined statistical-spectral approach can fully capture the physics of roughness-induced drag.

Furthermore, it is observed in figure~\ref{fig:LRP_filter} that the mean velocity profiles of original and filtered can exhibit some deviation very close to the wall. These deviations can be attributed to the altered volume occupied by roughness close to the wall as reflected by their $k_{\text{md}}$ and $Sk$ values. However, these
do not seem to have a significant influence beyond the region occupied by roughness. 
 
    \begin{figure}
    \centering
    \begin{tikzpicture}
\definecolor{kitgreen}{rgb}{0.00,0.58,0.51}
\definecolor{kitred}{rgb}{0.63,0.13,0.14}
\definecolor{kitblue}{rgb}{0.27,0.39,0.66}
\pgfplotsset{
    scale only axis,
}
\centering
    \begin{groupplot}[group style={group size= 2 by 3}]
            \nextgroupplot[
		ylabel={$k/\text{H}$},
		ylabel shift = 3.5 pt,
		xmin=0,xmax=1,
		ymin=0,ymax=0.3,
		unit vector ratio=1 1 1,
        title={(a) roughness A, original},
		clip=true,
		set layers,
		clip mode=individual,
		width=.4\linewidth,
		xtick={0,1},
		ytick={0,0.3},
        yticklabels={0,0.1},
		label style={font=\footnotesize},
		tick label style={font=\footnotesize},
	]
	\centering
\addplot [thick, color=blue, on layer=axis background]
graphics[xmin=-1,ymin=0,xmax=1,ymax=0.3]{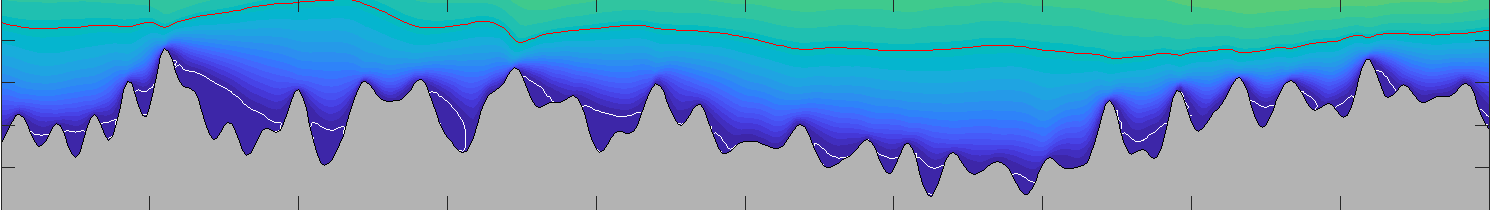};
\draw[<->] (0.219,0.1) -- (0.219,0.22);
\node [right,black] at (0.2,0.17) {\scriptsize $\Delta D_{\overline{u}^+=5}$};
    
             \coordinate (top) at (rel axis cs:0.3,1);

                        \nextgroupplot[
		ylabel shift = 3.5 pt,
		xmin=0,xmax=1,
		ymin=0,ymax=0.3,
		unit vector ratio=1 1 1,
        title={(b) roughness A, filtered},
		clip=true,
		set layers,
		clip mode=individual,
		width=.4\linewidth,
		xtick={0,1},
  ytick={\empty},
		label style={font=\footnotesize},
		tick label style={font=\footnotesize},
	]
	\centering
\addplot [thick, color=blue, on layer=axis background]
graphics[xmin=-1,ymin=0,xmax=1,ymax=0.3]{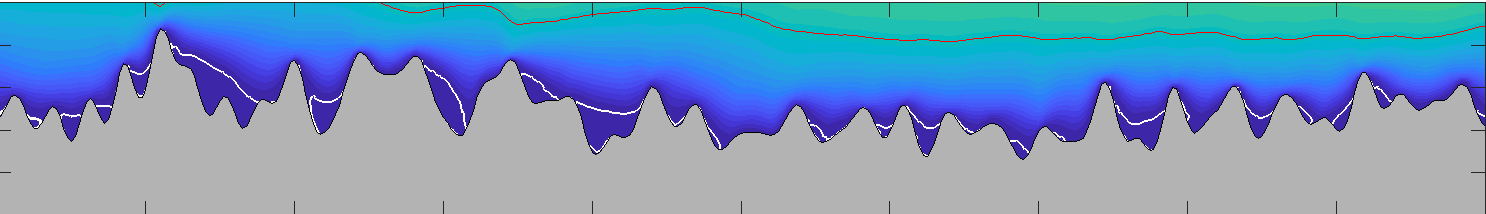};
         \nextgroupplot[
		ylabel={$k/\text{H}$},
		ylabel shift = 3.5 pt,
		xmin=1,xmax=2,
		ymin=0,ymax=0.3,
		unit vector ratio=1 1 1,
          title={(c) roughness B, original},
		clip=true,
		set layers,
		clip mode=individual,
		width=.4\linewidth,
		xtick={1,2},
    xticklabels={0,1},
		ytick={0,0.3},
        yticklabels={0,0.1},
		label style={font=\footnotesize},
		tick label style={font=\footnotesize},
	]
	\centering
\addplot [thick, color=blue, on layer=axis background]
graphics[xmin=0,ymin=0,xmax=2,ymax=0.3]{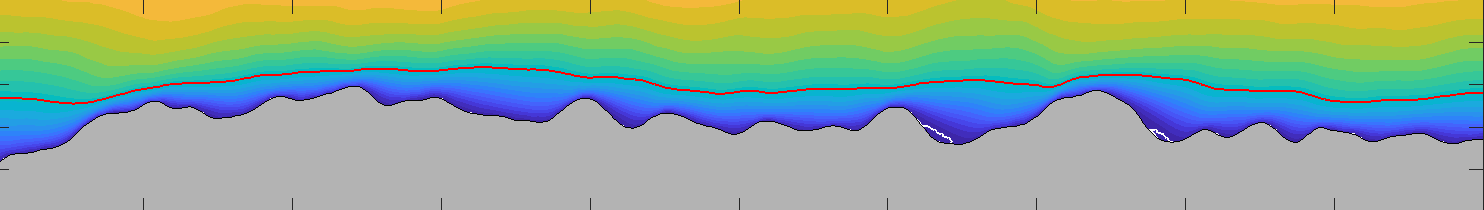};

 \nextgroupplot[
		ylabel shift = 3.5 pt,
		xmin=1,xmax=2,
		ymin=0,ymax=0.3,
		unit vector ratio=1 1 1,
		clip=true,
		set layers,
          title={(d) roughness B, filtered},
		clip mode=individual,
		width=.4\linewidth,
		xtick={1,2},
    xticklabels={0,1},
    ytick={\empty},
		label style={font=\footnotesize},
		tick label style={font=\footnotesize},
	]
	\centering
\addplot [thick, color=blue, on layer=axis background]
graphics[xmin=0,ymin=0,xmax=2,ymax=0.3]{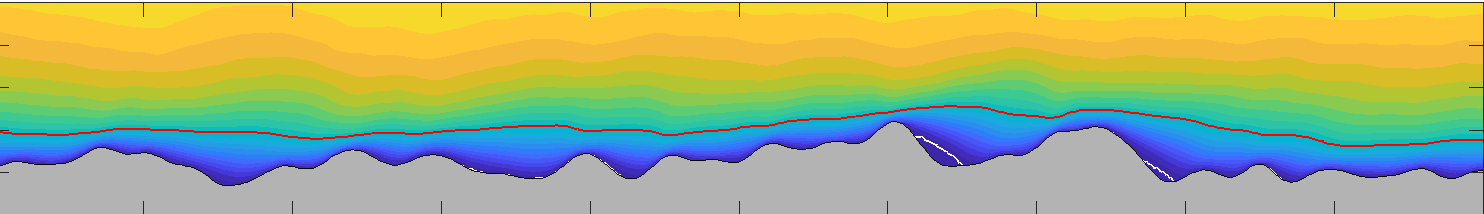};
         \nextgroupplot[
		xlabel={$x/\text{H}$},
				ylabel={$k/\text{H}$},
		ylabel shift = 3.5 pt,
		xmin=0,xmax=1,
		ymin=0,ymax=0.3,
		unit vector ratio=1 1 1,
  title={(e) roughness C, original},
		clip=true,
		set layers,
		clip mode=individual,
		width=.4\linewidth,
		xtick={0,1},
		ytick={0,0.3},
        yticklabels={0,0.1},
		label style={font=\footnotesize},
		tick label style={font=\footnotesize},
  colorbar,
		point meta min=0.0,
		point meta max=10,
				colorbar horizontal,
		colorbar style={
				xlabel={$\overline{u}^+$},
				xlabel shift = -6 pt,
		xtick={0,5,10},
		height=0.2cm,
    colorbar style={
    at={(rel axis cs:0,1.2)},anchor=north west}
		}
	]
	\centering
\addplot [thick, color=blue, on layer=axis background]
graphics[xmin=0,ymin=0,xmax=2,ymax=0.3]{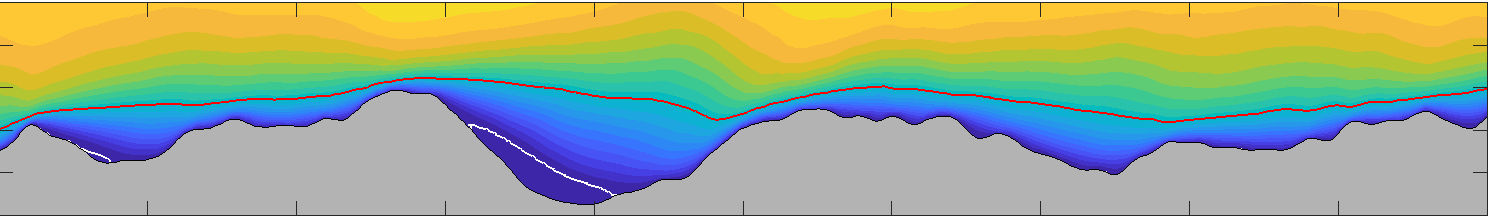};
\draw [->,color=black,thick] (0,-0.07) -- (0.2,-0.07);
\node [below] at (axis cs:0.1,-0.06) { \tiny \textbf{flow}};
 \nextgroupplot[
		xlabel={$x/\text{H}$},
		ylabel shift = 3.5 pt,
		xmin=0,xmax=1,
		ymin=0,ymax=0.3,
		unit vector ratio=1 1 1,
  title={(f) roughness C, filtered},
		clip=true,
		set layers,
		clip mode=individual,
		width=.4\linewidth,
		xtick={0,1},
    ytick={\empty},
		label style={font=\footnotesize},
		tick label style={font=\footnotesize},
  colorbar,
		point meta min=0.0,
		point meta max=10,
				colorbar horizontal,
		colorbar style={
				xlabel={$\overline{u}^+$},
				xlabel shift = -6 pt,
		xtick={0,5,10},
		height=0.2cm,
    colorbar style={
    at={(rel axis cs:0,1.2)},anchor=north west}
		}
	]
	\centering
\addplot [thick, color=blue, on layer=axis background]
graphics[xmin=0,ymin=0,xmax=2,ymax=0.3]{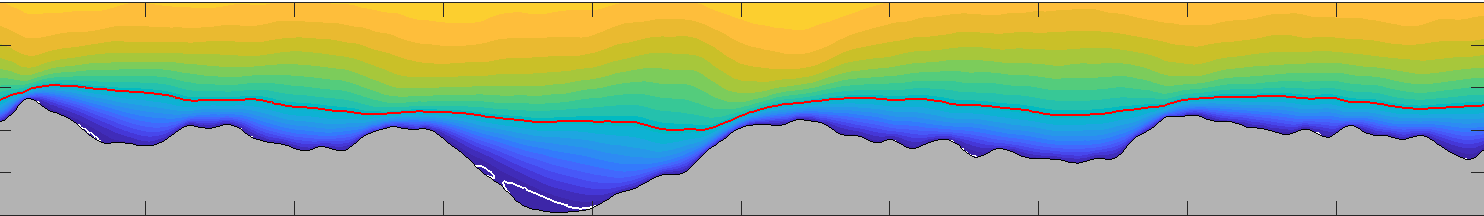};
\draw [->,color=black,thick] (0,-0.07) -- (0.2,-0.07);
\node [below] at (axis cs:0.1,-0.06) { \tiny \textbf{flow}};
             \coordinate (bot) at (rel axis cs:1,0);
    \end{groupplot}
    
\end{tikzpicture}
    \caption{Time averaged streamwise velocity distribution $\overline{u}^+$ in selected $z$-normal plane for the original and filtered cases A-C. The overlaid white contour lines mark the regions of reversed flow ($\overline{u}<0$). Blanketing layer (iso-contours of $\overline{u}^+=5$) is displayed with red contour lines. The gray color represents the rough structures. The calculation of blanking layer depth $\Delta D_{\overline{u}^+=5}$ is schematically illustrated in (a). }
    \label{fig:U2DV_RF}
\end{figure}
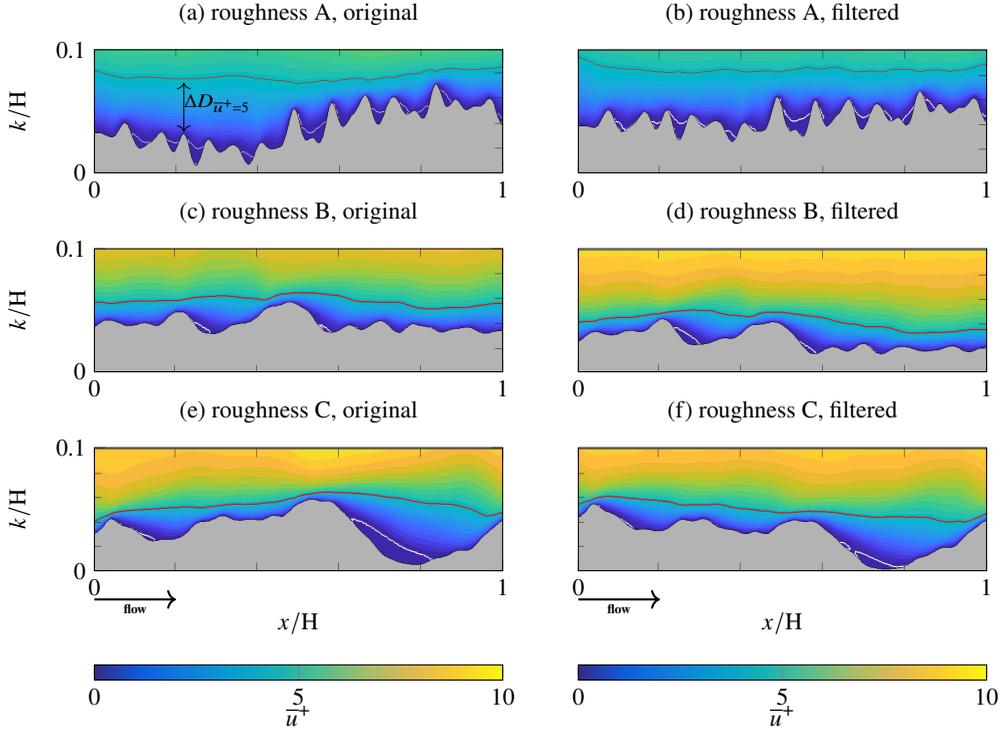
    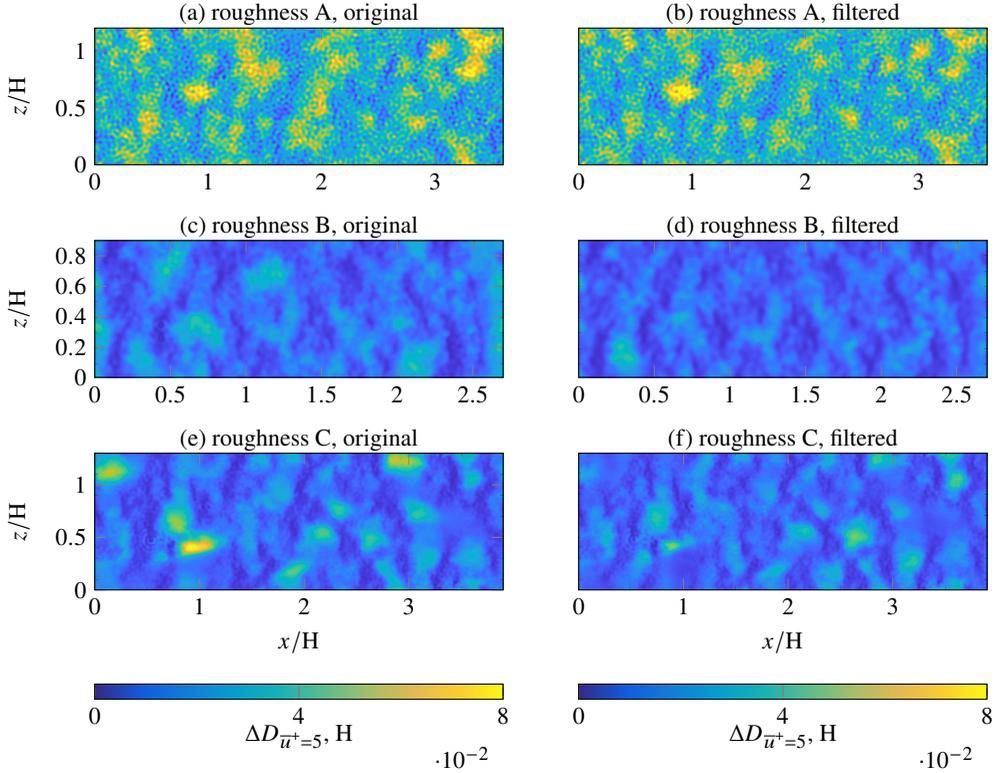
\begin{figure}
    \centering
    \begin{tikzpicture}
\definecolor{kitgreen}{rgb}{0.00,0.58,0.51}
\definecolor{kitred}{rgb}{0.63,0.13,0.14}
\definecolor{kitblue}{rgb}{0.27,0.39,0.66}
\pgfplotsset{
    scale only axis,
}
\centering
    \begin{groupplot}[group style={group size= 2 by 3}]
            \nextgroupplot[
		ylabel={$z/\text{H}$},
		ylabel shift = 3.5 pt,
		xmin=0,xmax=3.6,
		ymin=0,ymax=1.2,
		unit vector ratio=1 1 1,
		clip=true,
		set layers,
		clip mode=individual,
		width=.4\linewidth,
		label style={font=\footnotesize},
		tick label style={font=\footnotesize},
	]
	\centering
\addplot [thick, color=blue, on layer=axis background]
graphics[xmin=0,ymin=0,xmax=3.6,ymax=1.2]{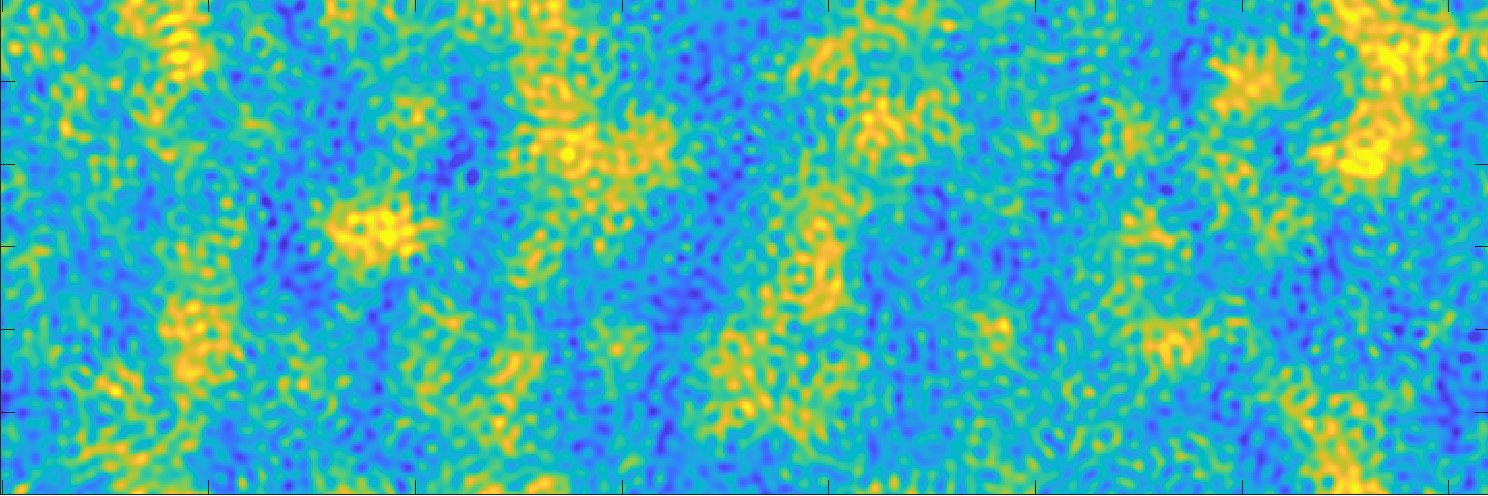};
            \node [black] at (rel axis cs: 0.5,1.1) { (a) roughness A, original};

             \coordinate (top) at (rel axis cs:0.3,1);

                        \nextgroupplot[
		ylabel shift = 3.5 pt,
		xmin=0,xmax=3.6,
		ymin=0,ymax=1.2,
		unit vector ratio=1 1 1,
		clip=true,
		set layers,
		clip mode=individual,
		width=.4\linewidth,
        ytick={\empty},
		label style={font=\footnotesize},
		tick label style={font=\footnotesize},
	]
	\centering
\addplot [thick, color=blue, on layer=axis background]
graphics[xmin=0,ymin=0,xmax=3.6,ymax=1.2]{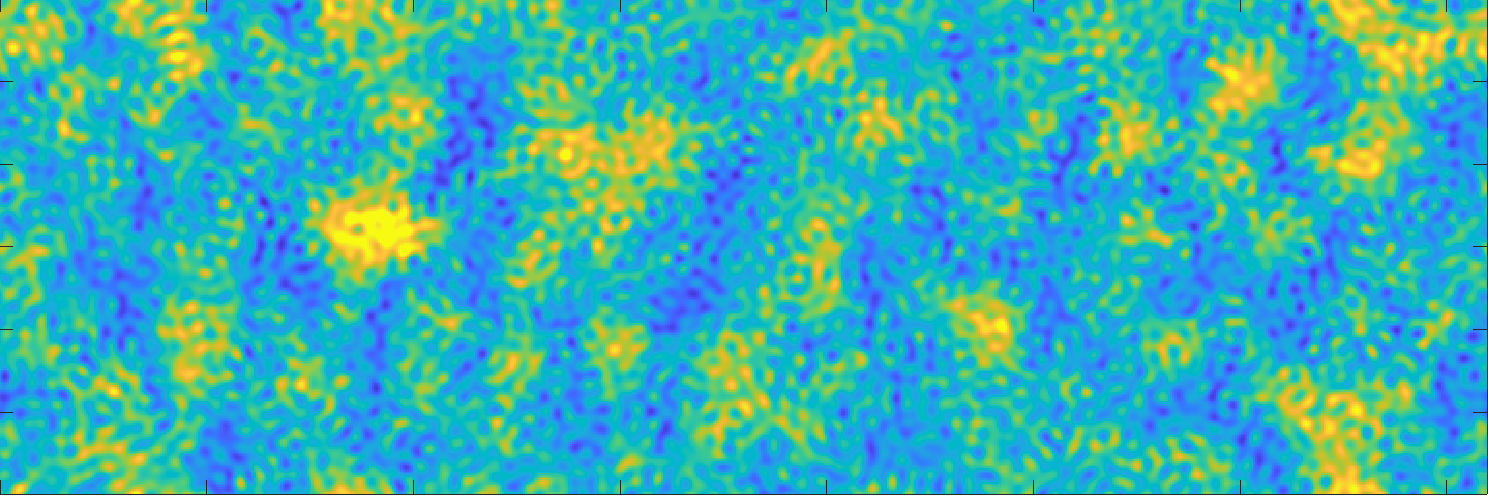};
            \node [black] at (rel axis cs: 0.5,1.1) {  (b) roughness A, filtered};

         \nextgroupplot[
		ylabel={$z/\text{H}$},
		ylabel shift = 3.5 pt,
		xmin=0,xmax=2.7,
		ymin=0,ymax=0.9,
		unit vector ratio=1 1 1,
		clip=true,
		set layers,
		clip mode=individual,
		width=.4\linewidth,
		label style={font=\footnotesize},
		tick label style={font=\footnotesize},
	]
	\centering
\addplot [thick, color=blue, on layer=axis background]
graphics[xmin=0,ymin=0,xmax=2.7,ymax=0.9]{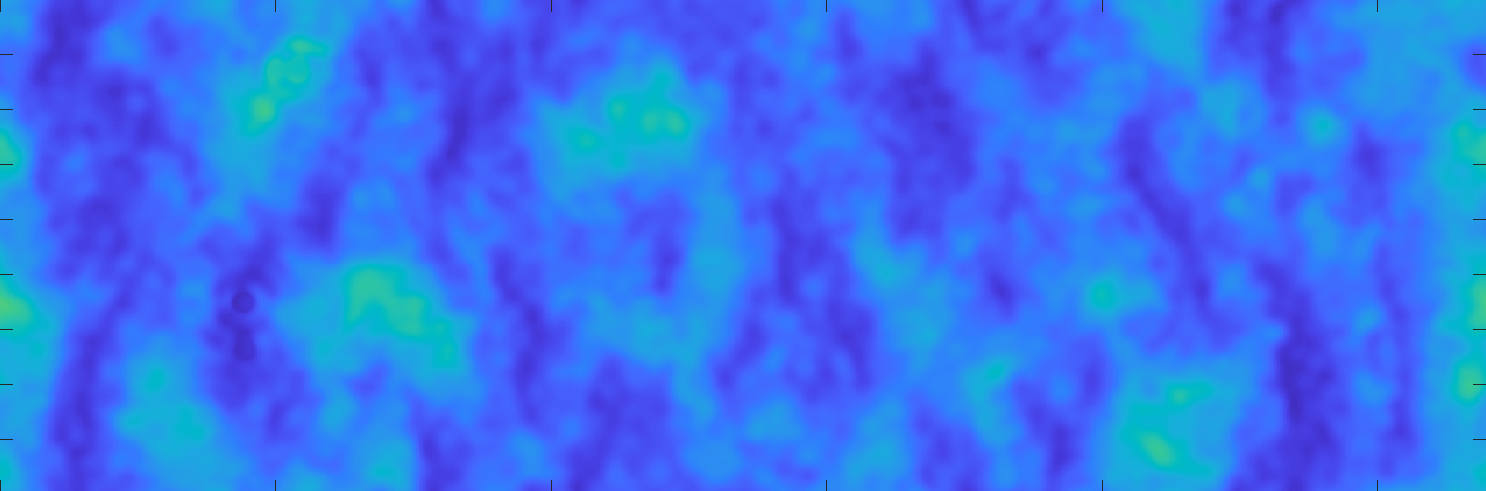};
            \node [black] at (rel axis cs: 0.5,1.1) {  (c) roughness B, original};

 \nextgroupplot[
		ylabel shift = 3.5 pt,
		xmin=0,xmax=2.7,
		ymin=0,ymax=0.9,
		unit vector ratio=1 1 1,
		clip=true,
		set layers,
		clip mode=individual,
		width=.4\linewidth,
    ytick={\empty},
		label style={font=\footnotesize},
		tick label style={font=\footnotesize},
	]
	\centering
\addplot [thick, color=blue, on layer=axis background]
graphics[xmin=0,ymin=0,xmax=2.7,ymax=0.9]{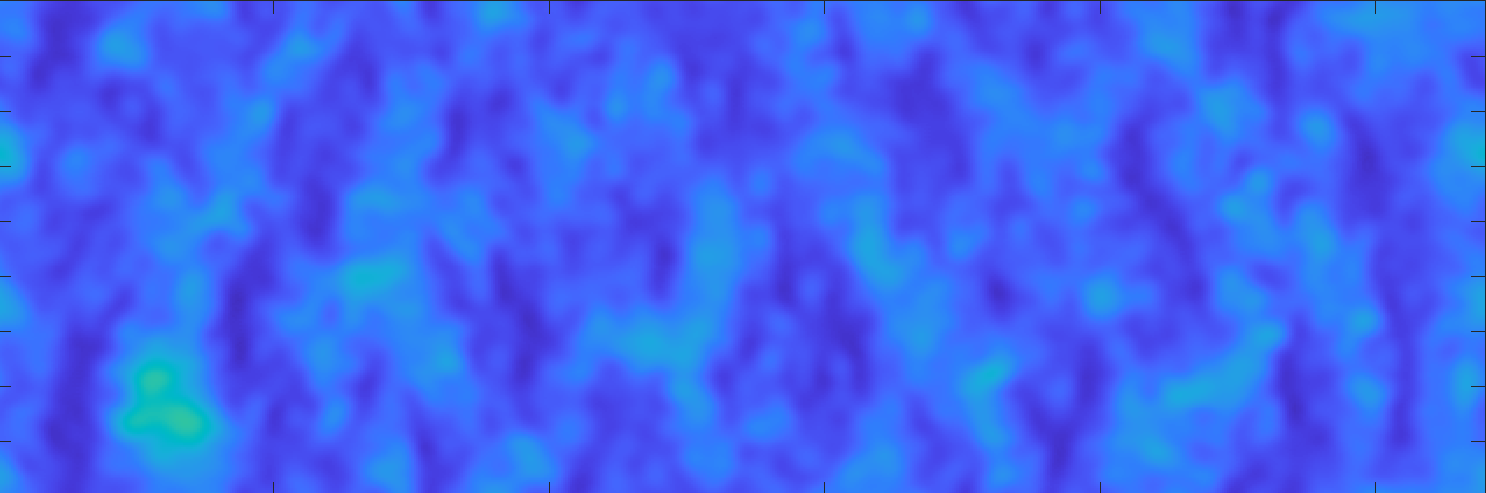};
            \node [black] at (rel axis cs: 0.5,1.1) {  (d) roughness B, filtered};

         \nextgroupplot[
		xlabel={$x/\text{H}$},
				ylabel={$z/\text{H}$},
		ylabel shift = 3.5 pt,
		xmin=0,xmax=3.9,
		ymin=0,ymax=1.3,
		unit vector ratio=1 1 1,
		clip=true,
		set layers,
		clip mode=individual,
		width=.4\linewidth,
		label style={font=\footnotesize},
		tick label style={font=\footnotesize},
  colorbar,
		point meta min=0.0,
		point meta max=0.08,
				colorbar horizontal,
		colorbar style={
				xlabel={$\Delta D_{\overline{u}^+=5}$, $\text{H}$},
				xlabel shift = -6 pt,
		xtick={0,0.04,0.08},
		height=0.2cm,
    colorbar style={
    at={(rel axis cs:0,1.2)},anchor=north west}
		}
	]
	\centering
\addplot [thick, color=blue, on layer=axis background]
graphics[xmin=0,ymin=0,xmax=3.9,ymax=1.3]{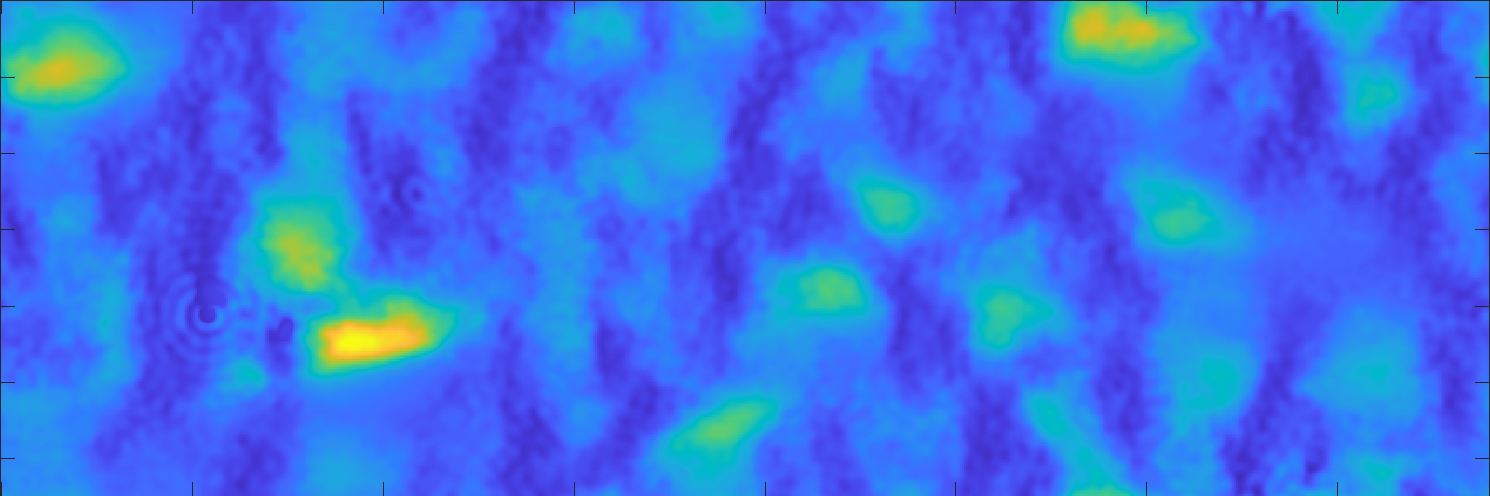};
            \node [black] at (rel axis cs: 0.5,1.1) {  (e) roughness C, original};

 \nextgroupplot[
		xlabel={$x/\text{H}$},
		ylabel shift = 3.5 pt,
		xmin=0,xmax=3.9,
		ymin=0,ymax=1.3,
		unit vector ratio=1 1 1,
		clip=true,
		set layers,
		clip mode=individual,
		width=.4\linewidth,
    ytick={\empty},
		label style={font=\footnotesize},
		tick label style={font=\footnotesize},
  colorbar,
		point meta min=0.0,
		point meta max=0.08,
				colorbar horizontal,
		colorbar style={
				xlabel={$\Delta D_{\overline{u}^+=5}$, $\text{H}$},
				xlabel shift = -6 pt,
		xtick={0,0.04,0.08},
  xticklabel style={
  /pgf/number format/precision=3,
  /pgf/number format/fixed},
		height=0.2cm,
    colorbar style={
    at={(rel axis cs:0,1.2)},anchor=north west}
		}
	]
	\centering
\addplot [thick, color=blue, on layer=axis background]
graphics[xmin=0,ymin=0,xmax=3.9,ymax=1.3]{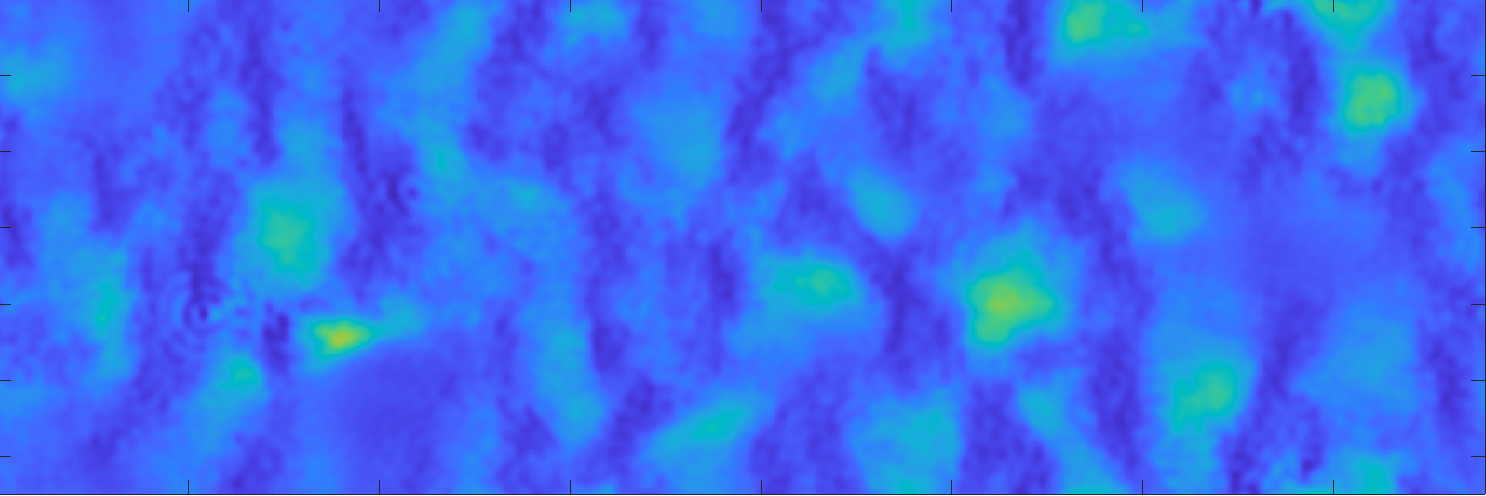};
            \node [black] at (rel axis cs: 0.5,1.1) {  (f) roughness C, filtered};
             \coordinate (bot) at (rel axis cs:1,0);
    \end{groupplot}
    
\end{tikzpicture}
    \caption{Blanketing layer depth $\Delta D_{\overline{u}^+=5}(x,z)=y_{\overline{u}^+=5}(x,z)-k(x,z)$ measured from rough surface for the original and filtered cases A-C. }
    \label{fig:Blanket}
\end{figure}

  {\color{black} 
To shed further light on why the filtered large scales do no contribute to added drag, exemplary $x$-$y$ planes of the time-averaged streamwise velocity field are examined in figure~\ref{fig:U2DV_RF}. The overlaid white contour lines are iso-contours of streamwise mean velocity $\overline{u}=0$, which mark the regions of reversed flow. Expectedly, roughness A exhibits relatively frequent flow recirculation due to larger local surface slope. In contrast, the occurrence of flow separation over roughness B and C seem to be less frequent, which could be linked to the waviness characteristics~\citep{Schultz2009} and less dominant form drag as a result of the low surface slope.  
When comparing the flow fields over filtered and original roughness, it is evident that the locations of flow recirculation are the same, and filtering has a minimal impact on the extent of reversed flow regions. 
Moreover, in figure~\ref{fig:U2DV_RF}, red contours are used to show the blanketing layer, which following~\citet{busse_thakkar_sandham_2017} is defined as the flow region confined by iso-surfaces of  $\overline{u}^+=5$. On a smooth wall, the blanketing layer would be identical to the viscous sub-layer while on a rough wall it can be an indication of how the near-wall flow adapts to the roughness topography. Similarly to the observations in~\citep{busse_thakkar_sandham_2017}, one can observe in figure~\ref{fig:U2DV_RF} that the blanketing layers in the shown cases do not follow the small roughness scales and steep roughness patterns. This behaviour can be better recognised if the `depth' of the blanketing layer, i.e. $\Delta D_{\overline{u}^+=5}(x,z)=y_{\overline{u}^+=5}(x,z)-k(x,z)$, is considered.

The maps of $\Delta D_{\overline{u}^+=5}(x,z)$ are shown in figure~\ref{fig:Blanket}, where a visual inspection reveals relative insensitivity of the blanketing layers to the smaller scales of roughness topography (which appear when the roughness height is subtracted from the ${\overline{u}^+=5}$ iso-contour height). Interestingly, in the same figure, a fair level of similarity is observed between the $\Delta D_{\overline{u}^+=5}(x,z)$ maps of the corresponding original and filtered cases. This can be a hint that the blanketing layer has adapted to the filtered scales. This idea is examined in appendix~\ref{appB} through a spectral analysis of $\Delta D_{\overline{u}^+=5}(x,z)$ for cases A-C. Based on this analysis, one might be able to hypothesize that the drag-irrelevant roughness scales are indeed those to which the blanketing layer can adapt.
As a final remark, a relation between the drag and blanketing layer depth is physically plausible as a change in this depth is generally accompanied by modifications in local flow phenomena (flow separation, strong changes in local velocity gradient on the wall, etc.) that can be linked to added drag. 

}

\subsection{Turbulent statistics over original and filtered roughness}
In the previous section, we used an LRP analysis of the trained model to identify which roughness scales contribute to the added skin-friction drag. 
While $\Delta U^+$ is arguably the most important flow statistic in practical sense, due to its relation to drag, roughness also affects higher order flow statistics particularly in the so-called `roughness sub-layer'~\citep{Chung2021annrev}. 
In the present study, we specifically focus on comparing the turbulent and dispersive stresses over pairs of unfiltered and filtered samples A, B, and C from section \ref{sec:LRP} as the main means of momentum transfer away from the wall. 

The velocity fluctuations in rough channels can be decomposed into turbulent and time-averaged spatial fluctuations following the triple decomposition of the velocity field proposed by~\citet{Raupach1992}:
\begin{equation}
    u_i(x,y,z,t)=\textlangle \overline{u}_i\textrangle(y) + \tilde{u}_i(x,y,z) + u^{\prime}_i(x,y,z,t)~.
\end{equation}
Here $\textlangle \overline{u}_i\textrangle(y)$ is the time- (overbar) and $x$-$z$-plane- (angled bracket) averaged velocity, $\tilde{u}_i(x,y,z)=\overline{u_i}(x,y,z)-\textlangle\overline{u_i}\textrangle(y)$ is the spatial variation of the time-averaged velocity and $u^{\prime}_i(x,y,z,t)$ is the space- and time-dependent turbulent fluctuation. 
Extrinsic plane-averaging is utilized in the present calculation of statistics, i.e. the solid regions are included in the averaging procedure with zero velocity (similar to, e.g.~\citet{yuan_piomelli_2014,stroh_schaefer_frohnapfel_forooghi_2020}).
Based on the above decomposition, local turbulent stresses $\overline{u^\prime_i u^\prime_j}(x,y,z)$ can be interpreted as measure of momentum transfer due to turbulent fluctuations. 
Analogous to the local turbulent stresses, one can define the dispersive stresses $\textlangle \tilde{u}_i\tilde{u}_j \textrangle(y)$ as the momentum transfer due to roughness-induced spatial fluctuations. 
Furthermore, double-averaged (DA) turbulent stresses are calculated through spatial averaging of the local turbulent stresses, i.e. $\textlangle\overline{u^\prime_i u^\prime_j}\textrangle(y)$.

\begin{figure}
    \centering
    \input{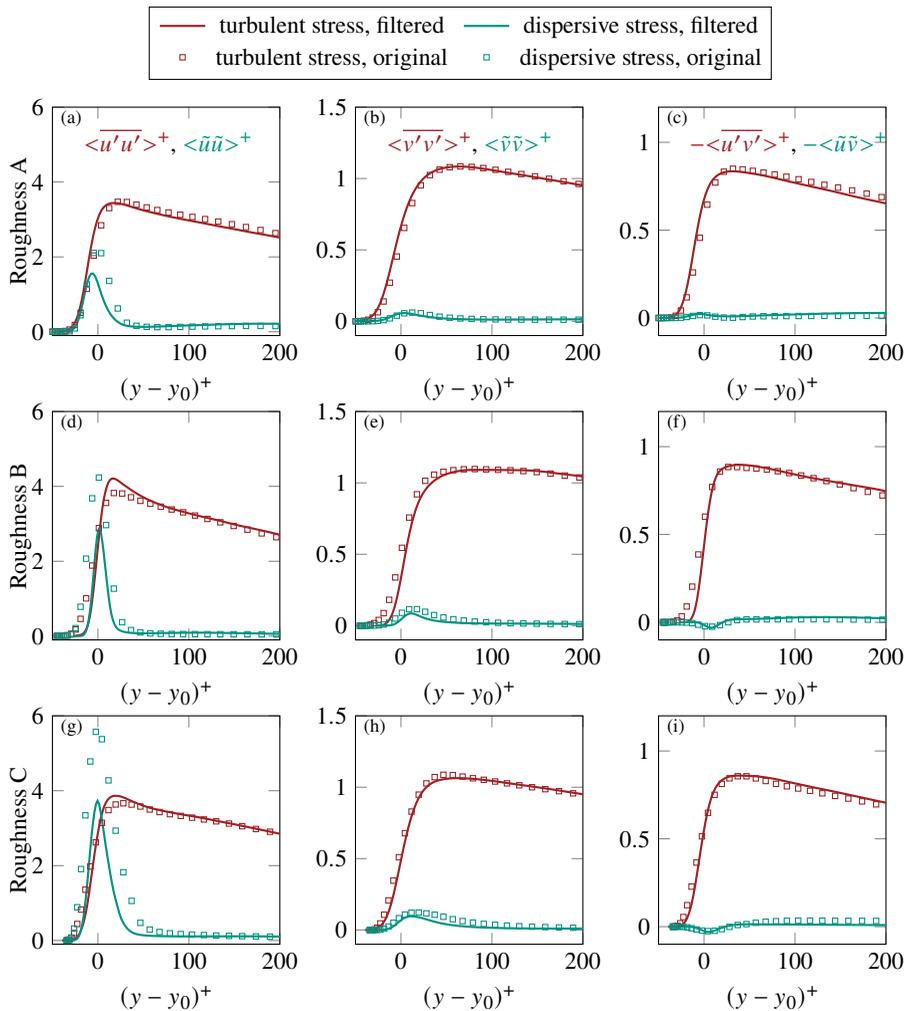}
    \caption{DA turbulent and dispersive stresses for Roughness A, B and C.}
    \label{fig:ReDis}
\end{figure}


The comparison of turbulent stresses for the three considered rough surfaces A, B, and C in filtered and unfiltered state are shown in the near wall region $(y-y_0)^+<200$ in figure~\ref{fig:ReDis} (red colour). 
Only a minor difference between original and filtered roughness can be observed for the turbulent Reynolds stresses. 
For the $\textlangle\overline{u^\prime u^\prime}\textrangle^+$ component, the peak values are comparable although, for samples B and C, filtering slightly increases the peak value. Roughness has previously been shown to damp inner-scaled streamwise turbulent stress that can be related to the suppression of elongated near-wall turbulent structures~\citep{yuan_piomelli_2014,PhysRevFluidsPourya}. This effect seems not to be affected significantly by elimination of drag-irrelevant large roughness scales.
Moreover, the agreement can be observed for the the other two normal turbulent stress as well as the shear stress ($\textlangle\overline{w^\prime w^\prime}\textrangle^+$ is not shown for the sake of brevity).
Excellent agreement of wall-normal turbulent stress is reminiscent of the suggestion by \citet{orlandi08} that roughness function is related to this component of turbulent stress.
Furthermore, the collapse of shear stress profiles is an indication of similarity in the vertical mean momentum transport due to turbulence. The agreement of these components thus contribute to the concordance of the mean velocity profiles in the log-layer.
For the dispersive stresses
it is apparent that the only component affected by filtering of roughness is the streamwise normal component $\textlangle\Tilde{u}\Tilde{u}\textrangle^+$, for which the peak values are reduced by filtering.
It is worth mentioning that same trend (reduction of the $\textlangle\Tilde{u}\Tilde{u}\textrangle^+$ peak values and agreement of other dispersive stresses) can be observed if an intrinsic averaging approach is used (not shown for brevity).
Despite the possible shift of the zero-plane $y_0$ after filtering, the peak of $\textlangle\Tilde{u}\Tilde{u}\textrangle^+$ is consistently observed at the vicinity of the respective zero-planes, i.e., at $(y-y_0)\approx0$. 
The discernible reduction in this peak value suggests a less pronounced inhomogeneity of the mean streamwise velocity when larger wavelengths are filtered. 
Arguably, the large-scale undulations present in the original roughness lead to large scale variations in mean velocity, resulting in greater flow inhomogeneity as also pointed out by~\citet{Yuan2018}. 

Despite the fact that the values of dispersive shear stress are small in all cases, a comparison among the three shown cases can provide certain insight into the the roughness-flow interactions. As depicted in figure~\ref{fig:ReDis} (c, f, i), roughness A exhibits a positive $-\textlangle\Tilde{u}\Tilde{v}\textrangle^+$ peak, whereas roughness B and C display negative peaks. Such a negative sign can be attributed to the `waviness effect' since a wavy structure (one with relatively low slope) causes an acceleration of the mean flow on the windward side and a deceleration on the leeward side \citep{Alvesportela21}. Positive $-\textlangle\Tilde{u}\Tilde{v}\textrangle^+$, on the other hand, can be linked to recirculation behind steep roughness elements \citep{Yuan2018}. This is in-line with the fact that roughness A has a much larger ES compared to the other two. 
The collapse of dispersive shear stress profiles in figure~\ref{fig:ReDis}~shows that none of these behaviours are affected by the applied filtering.

\begin{figure}
    \centering
    \begin{tikzpicture}
\begin{groupplot}[group style={group size= 3 by 2}]
        \nextgroupplot[
        title={Roughness A},
                ylabel shift = -6 pt,
        ylabel={$\Tilde{v}_{in}^+$},
        xlabel={$\Tilde{u}_{in}^+$},
		xmin=-6,xmax=6,
		ymin=-2,ymax=2,
		clip=true,
		set layers,
		clip mode=individual,
		width=.34\linewidth,
            height=.34\linewidth,
		ytick={-2.0,0,2.0},
		xtick={-6,-4,-2,0,2,4,6},
		label style={font=\footnotesize},
		tick label style={font=\footnotesize},
	]
	\centering
\addplot [black,thick] coordinates { (0,-2) (0,-2.1) };
\label{originalC}
\addplot [red,thick] coordinates { (0,-2) (0,-2.1) };
\label{filteredC}
\addplot [thick, color=blue, on layer=axis background]
graphics[xmin=-6,ymin=-2,xmax=6,ymax=2]{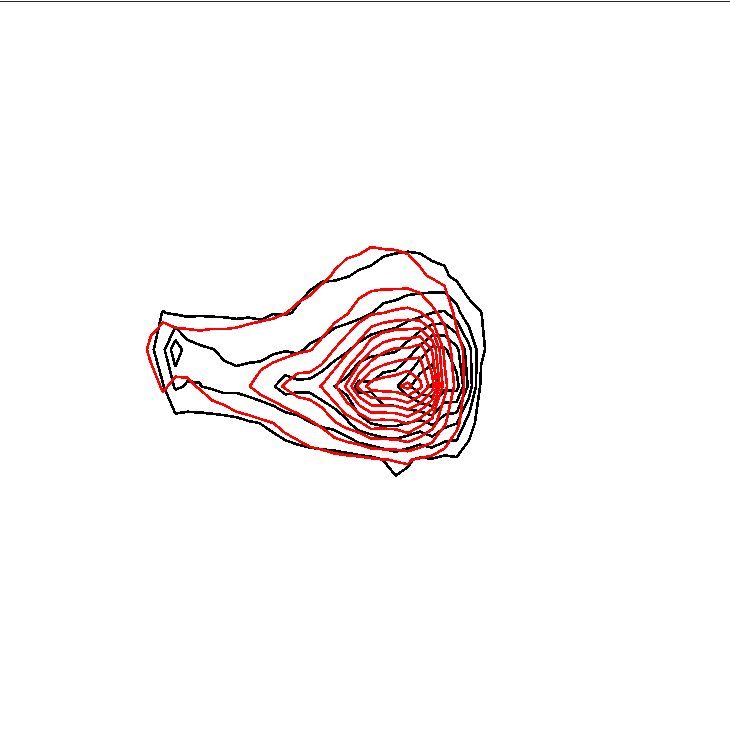};
\node[] at (4,1) {Q1};
\node[] at (-4,1) {Q2};
\node[] at (-4,-1) {Q3};
\node[] at (4,-1) {Q4};
\node[] at (-5,1.5) {(a)};
\draw[dashed,red,thick] (-6,0) -- (6,0);
\draw[dashed,red,thick] (0,-2) -- (0,2);

 \coordinate (top) at (rel axis cs:0.2,1);

        \nextgroupplot[
        xlabel={$\Tilde{u}_{in}^+$},
        title={Roughness B},
		xmin=-6,xmax=6,
		ymin=-2,ymax=2,
		clip=true,
		set layers,
		clip mode=individual,
		width=.34\linewidth,
            height=.34\linewidth,
		ytick={-2.0,0,2.0},
		xtick={-6,-4,-2,0,2,4,6},
		label style={font=\footnotesize},
		tick label style={font=\footnotesize},
	]
	\centering
\addplot [thick, color=blue, on layer=axis background]
graphics[xmin=-6,ymin=-2,xmax=6,ymax=2]{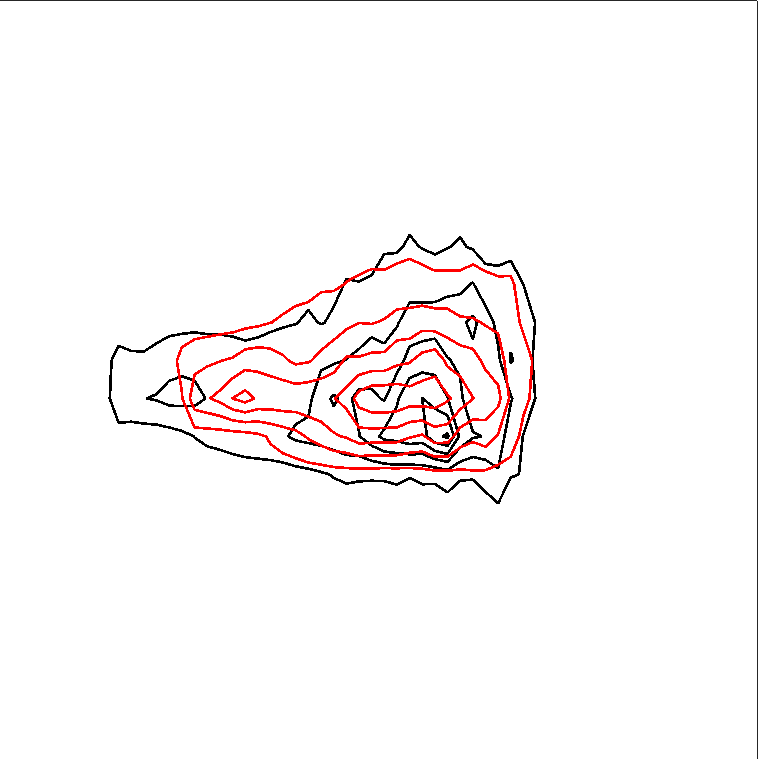};
\node[] at (4,1) {Q1};
\node[] at (-4,1) {Q2};
\node[] at (-4,-1) {Q3};
\node[] at (4,-1) {Q4};
\node[] at (-5,1.5) {(b)};
\draw[dashed,red,thick] (-6,0) -- (6,0);
\draw[dashed,red,thick] (0,-2) -- (0,2);
        \nextgroupplot[
        title={Roughness C},       
        xlabel={$\Tilde{u}_{in}^+$},
		xmin=-6,xmax=6,
		ymin=-2,ymax=2,
		clip=true,
		set layers,
		clip mode=individual,
		width=.34\linewidth,
            height=.34\linewidth,
		ytick={-2.0,0,2.0},
		xtick={-6,-4,-2,0,2,4,6},
		label style={font=\footnotesize},
		tick label style={font=\footnotesize},
	]
	\centering
\addplot [thick, color=blue, on layer=axis background]
graphics[xmin=-6,ymin=-2,xmax=6,ymax=2]{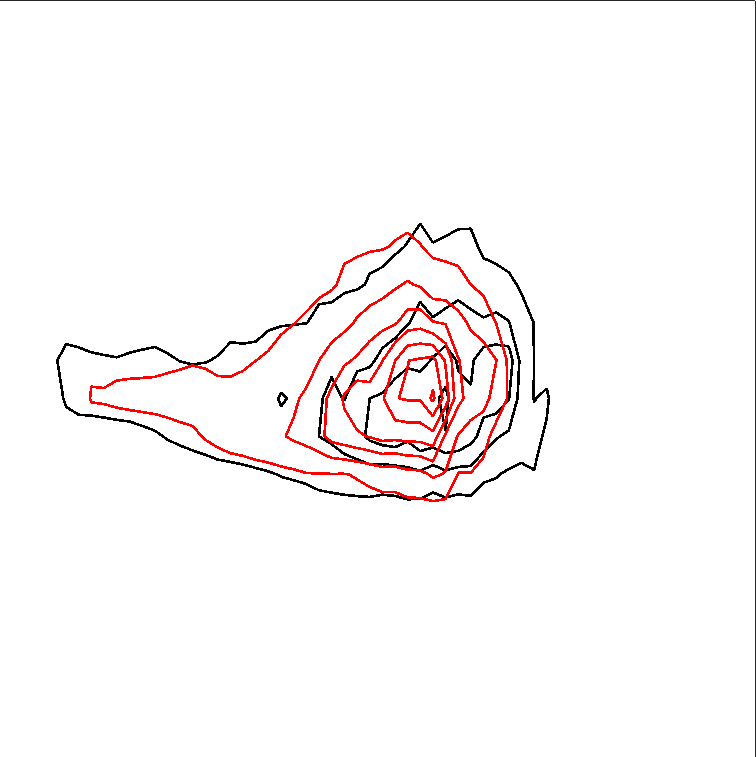};
\node[] at (4,1) {Q1};
\node[] at (-4,1) {Q2};
\node[] at (-4,-1) {Q3};
\node[] at (4,-1) {Q4};
\node[] at (-5,1.5) {(c)};
\draw[dashed,red,thick] (-6,0) -- (6,0);
\draw[dashed,red,thick] (0,-2) -- (0,2);
 \coordinate (bot) at (rel axis cs:1,0);
    \end{groupplot}
    
    \path (top-|current bounding box.west)-- 
          node[anchor=south,rotate=90]{}
          (bot-|current bounding box.west);
\path (top|-current bounding box.north)--
      coordinate(legendpos)
      (bot|-current bounding box.north);
\matrix[
    matrix of nodes,
    anchor=south,
    draw,
    inner sep=0.2em,
    draw
  ]at([yshift=1ex,xshift=-4ex]legendpos)
  {
    \ref{originalC}& \footnotesize{Original}&[2pt]
    \ref{filteredC}& \footnotesize{Filtered}\\};
\end{tikzpicture}
    \caption{Joint p.d.f. of $\Tilde{u}_{in}^+$ and $\Tilde{v}_{in}^+$ at plane $y=y_0$, values in roughness excluded. Contour lines range from 0.05 to 1.55 with step 0.1. Subscript $in$ indicates being a result of intrinsic averaging.}
    \label{fig:JPDF}
\end{figure}
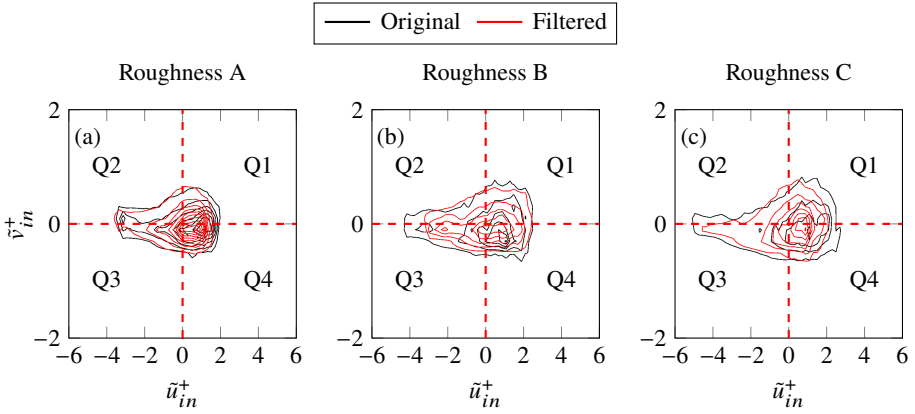

The results shown so far indicate that the streamwise dispersive stress is the only second-order one-point velocity statistic affected by filtering of drag-irrelevant scales. This however does not modify the shear stress profile as discussed above. To further elaborate this finding, joint p.d.f.s of local dispersive motions in the wall-parallel plane $y=y_0$ are calculated for all the three samples along with their filtered counterparts and shown in figure~\ref{fig:JPDF}. 
Here intrinsive averaging is used meaning that the areas inside roughness are excluded for calculation of the dispersive velocities shown in the joint p.d.f..
The subscript ${in}$ denotes intrinsic averaging. 
Following the idea of quadrant analysis~\citep{wallace_eckelmann_brodkey_1972}, the $\Tilde{u}_{in}^+$--$\Tilde{v}_{in}^+$ plane is divided into four quadrants, Q1-Q4, based on the signs of $\Tilde{u}_{in}^+$ and $\Tilde{v}_{in}^+$. While the joint p.d.f. look relatively similar before and after high-pass filtering, it is observed that filtering results in contours shrinking along the $\Tilde{u}_{in}^+$-axis. This is in-line with the reduction of peak values of streamwise dispersive component discussed before. 
Notably, the joint p.d.f. retains its near-symmetry with respect to the $\Tilde{v}_{in}^+$-axis, which means that reduction of extreme $\Tilde{u}_{in}^+$ fluctuations shows no preference in the direction of momentum transfer. This results in the similar shape of the contours apart from horizontal stretching. An obvious outcome is that the shear stress profiles are unaffected by modifications in $\Tilde{u}_{in}^+$. 

\section{Conclusions}
\label{sec:Conclu}
In this study, we present a new approach for predicting the normalized equivalent sand-grain height $k_r=k_s/k_{99}$ of homogeneous irregular roughness based on roughness p.d.f. and PS utilizing a machine learning ENN model.
The model is developed within the AL framework to effectively reduce the required amount of training data. This framework 
searches for roughness samples with the highest prediction variances $\sigma_{k_r}$ in an unlabeled repository $\mathcal{U}$ of 4200 samples.
Eventually, a labeled data set $\mathcal{L}$ comprising 85 AL-selected samples is constricted and utilized to derive the ENN model. 
The significant improvement of the learning efficiency of the model through AL is demonstrated by comparing it with a non-AL approach. Furthermore, it is observed that the employment of AL serves to effectively mitigate the deleterious effects of over-fitting, as evidenced by the observed general drop in prediction error. The mean prediction errors of the final AL-ENN model for an internal testing data set $\mathcal{T}_{\text{inter}}$, as well as two external data sets containing both realistic and artificially generated roughness, $\mathcal{T}_{\text{ext, 1}}$ and $\mathcal{T}_{\text{ext, 2}}$, are {\color{black} 9.3\%}, 5.2\%, and 10.2\%, respectively. The consistently good predictions for testing data with different natures can be taken as a sign that a universal model is approached.

Moreover, novel physical insights on the interactions between roughness and turbulent flow are sought by exploring the information embedded in the data-driven model.
To this end, the LRP technique is employed to evaluate the contributions of different wave-numbers in the discretized roughness PS towards the predicted value $k_r$.
The PS content identified with positive contribution according to the LRP are interpreted as `drag-relevant'. Subsequently, high-pass filtering is used to exclude the drag-irrelevant scales, and based on the DNS results for exemplary cases, it is observed that despite the considerable variations in the roughness appearance and statistics, the mean velocity profiles of these high-pass filtered samples collapse well into the original samples in the logarithmic layer, thus having the same $k_s$ values.
{\color{black} 
The LRP-identified drag-irrelevant structures are further studied through an analysis of the behaviour of the blanketing layers over filtered and original roughness. 
Similarity is observed when maps of blanketing layer `depth' $\Delta D_{\overline{u}^+=5}$ of filtered and original roughness are compared. This can indicate that the blanketing layer can adapt to the drag-irrelevant scales.
}
Furthermore, turbulent and dispersive stresses over original and filtered roughness are compared; it is shown that the turbulent stresses are not affected meaningfully by removal of the drag-irrelevant structures.
Agreement is observed for both turbulent and dispersive shear stresses which indicates identical momentum transport pattern in wall-normal direction over original and filtered roughness.
The sole effect of filtering observed on one-point second-order velocity statistics is the the reduced streamwise dispersive stress, which can be an indication of less inhomogeneity of mean flow over the filtered roughness. 
Finally, the joint p.d.f. of streamwise and wall-normal dispersive velocity components are compared for original and filtered roughness , and it is observed that the probability contours are generally similar, with the streamwise component having a smaller extent in the filtered case. No strong change of preference towards a certain quadrant results from filtering. 

In summary, according to the present results, it can be stated that use of roughness height p.d.f and PS as the model inputs, combined with an AL framework for exploring the vast parameter space, has the potential for developing universal roughness predictive models. Additionally, the present work shows a clear potential to extract physical information from the data-driven models through interpretation techniques, here the LRP. 

The LRP-based analysis presented in this work is obviously a first step towards utilizing data-driven models beyond merely predictive tools in the context of rough-wall turbulence. Future investigations can explore other avenues to extract knowledge on the roughness-turbulence interactions from such models. The present LRP-based analysis can also be further investigated towards more rigorous criteria for identifying the drag-irrelevant structures. Furthermore, the LRP-based filtering can be a basis for developing more accurate empirical correlations solely incorporating the drag-relevant structures. Finally, one should note that the present work merely focuses on roughness topographies of isotropic and homogeneous nature. Extension towards more general anisotropic and/or heterogeneous roughness is another obvious direction for future research.

\appendix
\section{Exemplary roughness at each round}\label{appA}
\begin{figure}
    \centering
    \includegraphics[width=\linewidth]{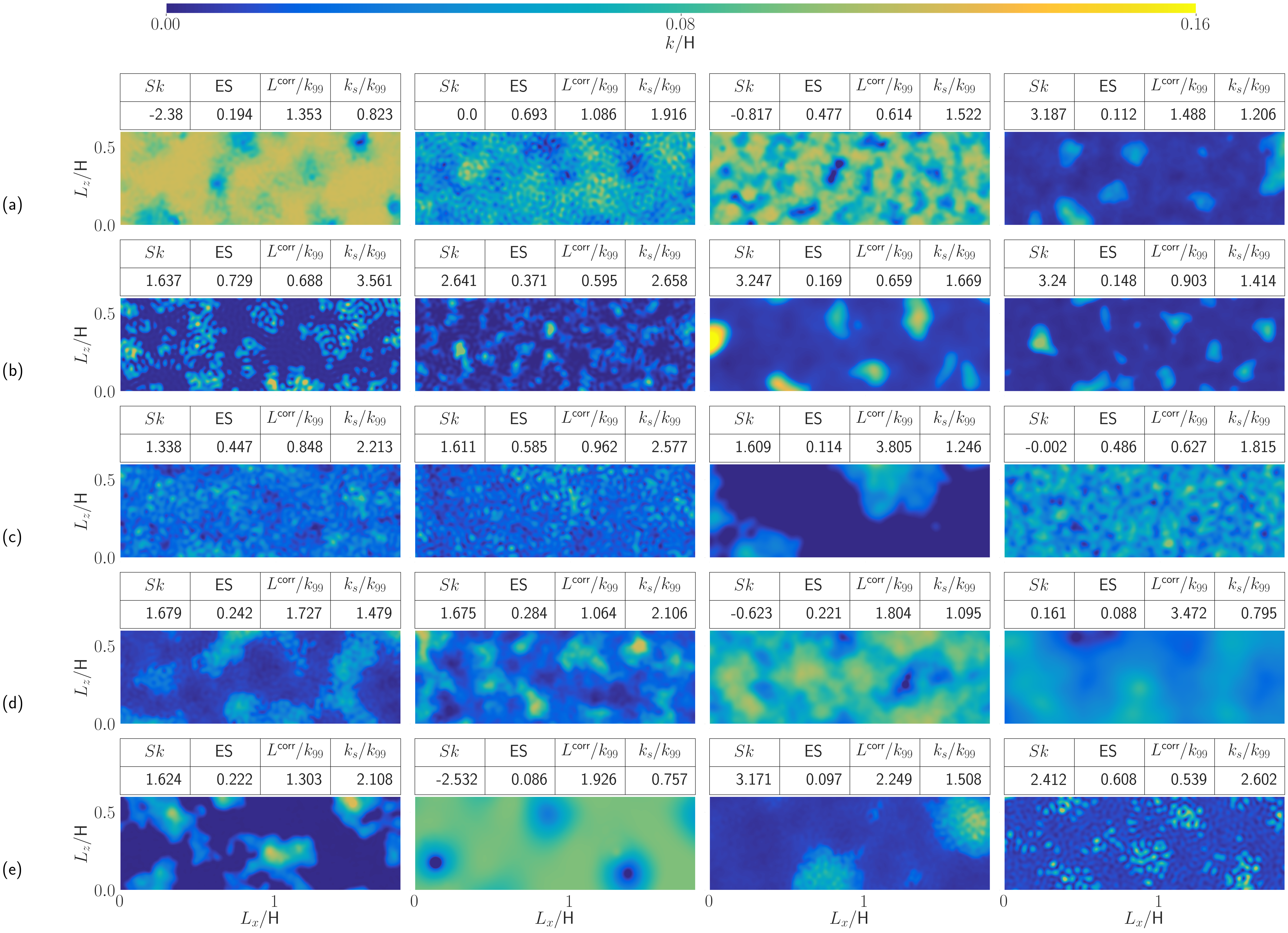}
     \caption{Examples of roughness samples included in $\mathcal{L}$. Patches of same size extracted from different samples. (a$\sim$e) correspond to initial round and AL round 1$\sim$4, respectively.}
    \label{fig:Roughness4Rounds}
    \end{figure}
Exemplary rough surfaces selected from each iteration along with their statistical parameters and $k_s$ values are shown in figure~\ref{fig:Roughness4Rounds}.
\section{Spectral analysis of blanketing layer depth}\label{appB}
\begin{figure}
    \centering
    \begin{tikzpicture}
    \begin{groupplot}[group style={group size= 3 by 1}]
        \nextgroupplot[
        			xlabel=$2\pi k_{99}/\lambda$,
		ylabel={Pre-multiplied spectra},
		xmin=.0549,xmax=11.9381,
		ymin=0,ymax=2,
		clip=true,
		set layers,
		xmode=log,
		ymode=linear,
		clip mode=individual,
		height=.3\textwidth,
		width=.3\textwidth,
        tick label style={font=\footnotesize},
                    legend style={font=\tiny,anchor=south west},
                        legend pos=south west,
	]
	  \addplot [
            kit-green100,thick
            ]
            coordinates{
            (100,2)			
			(100,1)
			};
			\label{Blanket_PS}
  \addplot [
            kit-red100,thick
            ]
            coordinates{
            (100,2)			
			(100,1)
			};
			\label{roughness_PS}
			              \addplot [
            gray,mark=star,only marks
            ]
            coordinates{
            (100,2)			
			(100,1)
			};
			\label{Blanket_PS_raw}
              \addplot [
            red,dashed,thick
            ]
            coordinates{
            (100,2)			
			(100,1)
			};
			\label{LRPFILTER}
\addplot [thick, color=blue, on layer=axis background]
graphics[xmin=.0549,ymin=0,xmax=11.9381,ymax=2]{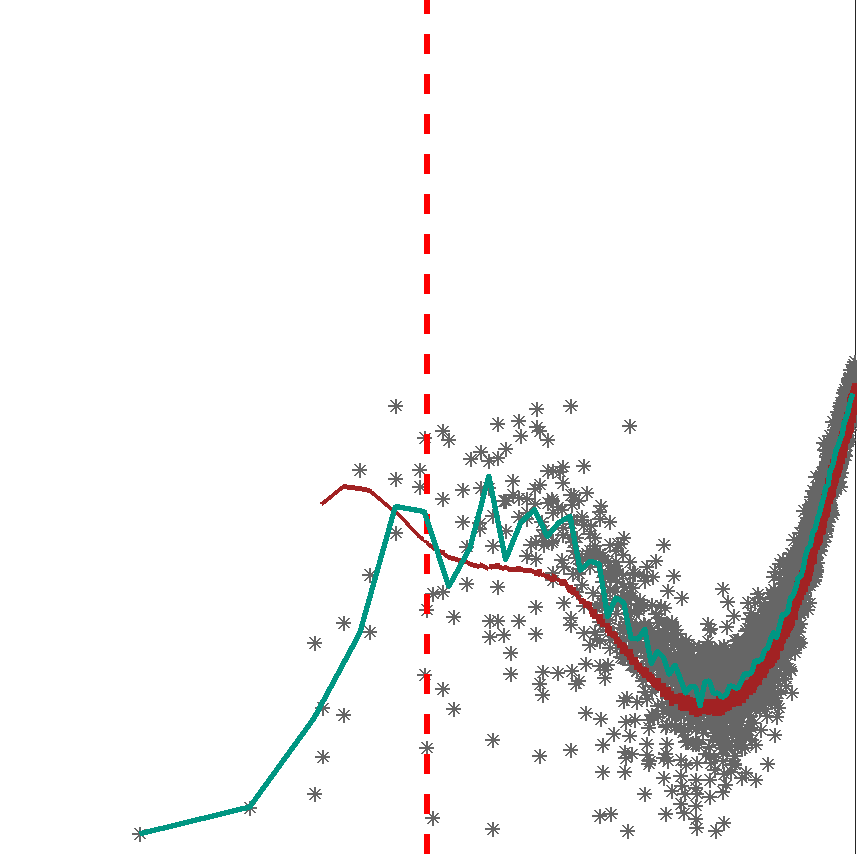};

\node[above right] at (current axis.north west)
{\scriptsize{(a) Roughness A}};
 \coordinate (top) at (rel axis cs:0,1);

\nextgroupplot[
        			xlabel=$2\pi k_{99}/\lambda$,
		xmin=.0654,xmax=10.8385,
		ymin=0,ymax=2,
		clip=true,
		set layers,
		xmode=log,
		ymode=linear,
		clip mode=individual,
		height=.3\textwidth,
		width=.3\textwidth,
        tick label style={font=\footnotesize},
                    legend style={font=\tiny,anchor=south west},
                        legend pos=south west,
	]
\addplot [thick, color=blue, on layer=axis background]
graphics[xmin=.0654,ymin=0,xmax=10.8385,ymax=2]{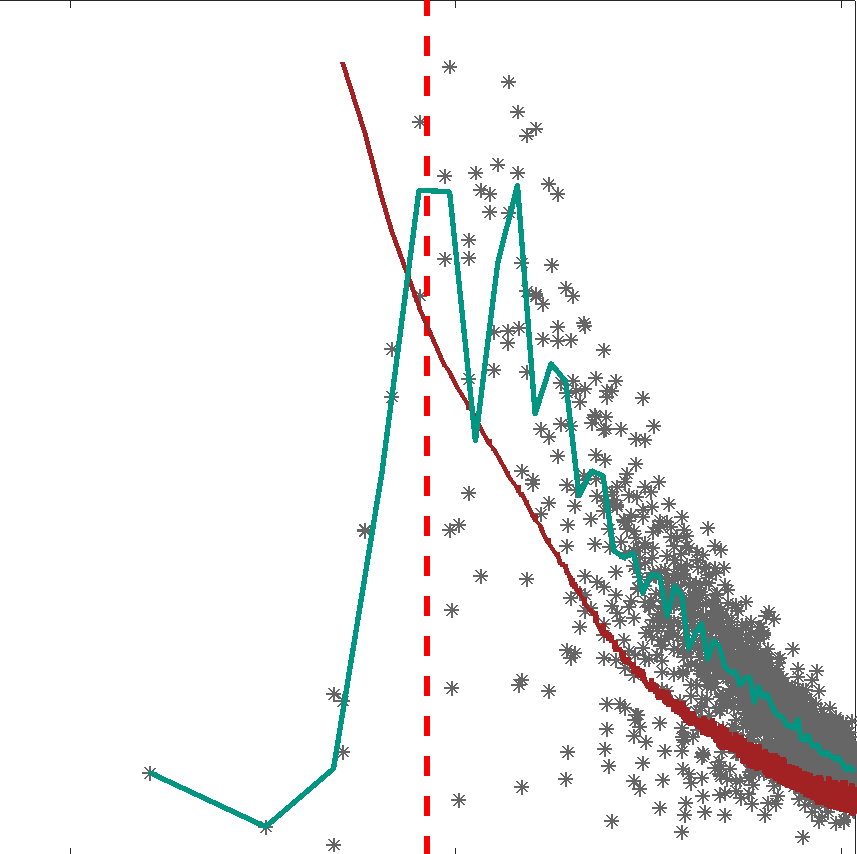};

\node[above right] at (current axis.north west)
{\scriptsize{(b) Roughness B}};

\nextgroupplot[
        			xlabel=$2\pi k_{99}/\lambda$,
		xmin=.0389,xmax=12.2522,
		ymin=0,ymax=2,
		clip=true,
		set layers,
		xmode=log,
		ymode=linear,
		clip mode=individual,
		height=.3\textwidth,
		width=.3\textwidth,
        tick label style={font=\footnotesize},
                    legend style={font=\tiny,anchor=south west},
                        legend pos=south west,
	]
\addplot [thick, color=blue, on layer=axis background]
graphics[xmin=.0389,ymin=0,xmax=12.2522,ymax=2]{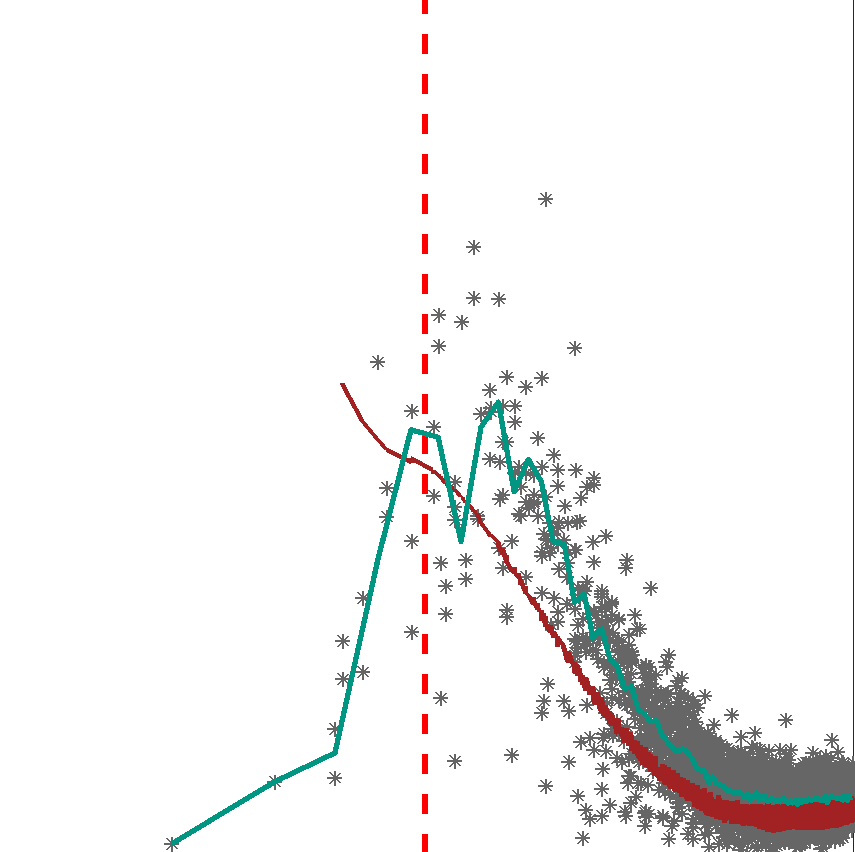};

\node[above right] at (current axis.north west)
{\scriptsize{(c) Roughness C}};
            \coordinate (bot) at (rel axis cs:1,0);
    \end{groupplot}
    \path (top-|current bounding box.west)-- 
          node[anchor=south,rotate=90]{}
          (bot-|current bounding box.west);
\path (top|-current bounding box.north)--
      coordinate(legendpos)
      (bot|-current bounding box.north);
\matrix[
    matrix of nodes,
    anchor=south,
    draw,
    inner sep=0.2em,
    draw
  ]at([yshift=1ex]legendpos)
  {
    \ref{roughness_PS}& \footnotesize Roughness&[2pt]
    \ref{Blanket_PS}& \footnotesize Blanketing depth, averaged&[2pt]
    \ref{Blanket_PS_raw}& \footnotesize Blanketing depth, raw&[2pt]
    \ref{LRPFILTER}& \footnotesize LRP-filter\\};
\end{tikzpicture}
    \caption{Pre-multiplied spectra of blanketing layer depth $\Delta D_{\overline{u}^+=5}$ overlaid with that of the corresponding roughness topography. Symbols indicate the spectrum in different directions while green line show the azimuthal average. The scatter of the symbols indicates the anisotropic characteristics of the map. Structures smaller than the smallest in-plane roughness wavelength $\lambda_1$ are omitted.}
    \label{fig:Blanket_PS}
\end{figure}
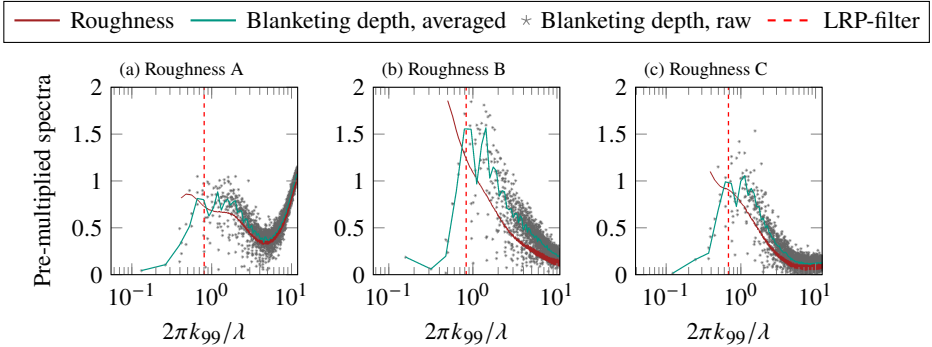
{\color{black}In order to further investigate possible links between the blanketing layer depth $\Delta D_{\overline{u}^+=5}(x,z)=y_{\overline{u}^+=5}(x,z)-k(x,z)$ and drag-irrelevant scales, we plot pre-multiplied PS of the blanketing layer depth maps over the original rough surfaces in figure~\ref{fig:Blanket_PS}. For more clarity, the two-dimensional spectra are azimuthally averaged and plotted in green color. Moreover, the locations of LRP-identified filters and the pre-multiplied roughness PSs are also added to the plots. Interestingly, all spectra show a significant decrease in the contribution of wavelengths larger than (wavenumbers smaller than) the filter. Note that all plots in figure~\ref{fig:Blanket_PS} belong to the original cases with no influence from the filtering. A wavelength that is present in the roughness topography but absent in the blanketing layer depth is one to which the layer has adapted. Therefore, the fact that the spectrum drops for drag-irrelevant scales might suggest that those scales are the ones to which the blanketing layer can adapt.

Despite the above discussion, further systematic investigations are required to establish a conclusive evidence as the present study covers limited ranges of roughness scales and Reynolds numbers. Ideally data on surfaces with more `drag-irrelevant' large scales and at a much wider range of Reynolds numbers are required to establish a solid hypothesis.
Additionally, one should bear in mind that the current results are obtained in the fully-rough regime, and as discussed by~\citet{busse_thakkar_sandham_2017}, blanketing layer can behave differently at different regimes.}
\backsection[Acknowledgements]{J. Yang gratefully acknowledges partial financial support from Friedrich und Elisabeth Boysen-Foundation (BOY-151). P. Forooghi gratefully acknowledges financial support from Aarhus Univesity Research Foundation (starting grant AUFF-F-2020-7-9). S. Lee and S. Bagheri sincerely appreciate financial support by Swedish Energy Agency under grant number 51554-1. This work was performed on the supercomputer HoreKa and the storage facility LSDF funded by the Ministry of Science, Research and the Arts Baden-Württemberg and by the Federal Ministry of Education and Research.}

\backsection[Declaration of interests]{The authors report no conflict of interest.}

\backsection[Data availability statement]{The trained model for prediction of $k_s$ can be publicly accessed at the project web-page, \url{http://roughness.org}, for research purposes  (website under construction at the time of submission). The generated database including labeled roughness samples is made publicly available through \url{http://roughnessdatabase.org}. The codes for roughness generation, statistical analysis, and model training can be downloaded from the first author's GitHub repository \url{https://github.com/JiashengY/Active-learning-codes} (code will be provided after the paper is accepted).}

\backsection[Author ORCID]{\\
J. Yang, https://orcid.org/0000-0003-0091-6855; \\
A. Stroh, https://orcid.org/0000-0003-0850-9883;\\
S. Lee, https://orcid.org/0000-0001-7341-8289; \\
S. Bagheri, https://orcid.org/0000-0002-8209-1449;\\
B. Frohnapfel, https://orcid.org/0000-0002-0594-7178;\\
P. Forooghi, https://orcid.org/0000-0001-9212-514X
}

\backsection[Author contributions]{\\
JY: methodology, investigation, data curation, software, formal analysis, visualization, writing – original draft\\
AS: methodology, data curation, formal analysis, supervision, writing – review and editing\\
SL: ML methodology, writing – review and editing\\
SB: ML methodology, formal analysis, writing – review and editing\\
BF: formal analysis, supervision, resources, funding acquisition, writing – review and editing\\
PF: conceptualization, methodology, formal analysis, supervision, writing – original draft, writing – review and editing, funding acquisition, project administration

}

\bibliographystyle{jfm}
\bibliography{jfm}

\end{document}